\documentclass[11pt]{amsart}

\usepackage{graphicx,euscript}
\usepackage{amssymb,url}

\usepackage{graphicx,euscript}
\usepackage{amsmath}
\usepackage{amsthm}
\usepackage{amssymb}
\usepackage{multirow,url,color}
\theoremstyle{plain}
\newtheorem{Thm}{Theorem}[section]
\newtheorem{Cor}[Thm]{Corollary}
\newtheorem{Lem}[Thm]{Lemma}
\newtheorem{Prop}[Thm]{Proposition}

\theoremstyle{definition}

\newcommand{\wt}{\widetilde}

\newcommand{\E}{\mathbb{E}}

\newcommand{\ls}{\bigl\{}
\newcommand{\rs}{\bigr\}}

\newcommand{\lp}{\bigl(}
\newcommand{\rp}{\bigr)}

\newcommand{\gs}{\sigma}

\newcommand{\gd}{\delta}
\newcommand{\gl}{\lambda}
\newcommand{\oh}{\frac{1}{2}}

\newcommand{\lnf}{\lim_{n \to \infty}}
\newcommand{\ltf}{\lim_{t \to \infty}}

\newcommand{\gep}{\epsilon}

\newcommand{\nc}{\newcommand}

\newcommand{\eu}{\EuScript}
\newcommand{\indic}{\boldsymbol{1}}

\newcommand{\on}{\operatorname}
\nc{\G}{\eu{G}}
\nc{\lip}{\on{Lip}}
\nc{\izf}{\int_0^\infty}
\nc{\tand}{\text{ and }}
\nc{\tst}{\text{ s.t. }}
\nc{\fM}{\mathfrak{M}}
\nc{\fP}{\mathfrak{P}}

\newcommand{\lb}{\bigl[}
\newcommand{\rb}{\bigr]}
\newcommand{\lv}{\bigl|}
\newcommand{\rv}{\bigr|}
\newcommand{\rec}{\frac{1}}

\newcommand{\R}{\mathbb{R}}

\nc{\N}{\mathbb{N}}
\nc{\mL}{\eu{L}}
\nc{\mA}{\eu{A}}
\nc{\M}{\eu{M}}
\nc{\C}{\eu{C}}
\nc{\B}{\eu{B}}

\nc{\vx}{\vec{x}}
\nc{\vy}{\vec{y}}
\nc{\DF}{\eu{F}}
\nc{\df}{f}
\nc{\tX}{\wt{X}}
\nc{\mE}{\mathbb{E}}
\nc{\brM}{\bar{\mM}}
\nc{\tih}{\tilde{h}}
\nc{\lep}{\frac{\gl}{\gep}}
\nc{\tp}{\tau_{\partial}}
\nc{\sM}{\mM^{*}}
\nc{\ns}{\nu^*}
\nc{\bP}{\bar{P}}
\nc{\bS}{\bar{S}}
\renewcommand{\P}{\mathbb{P}}
\nc{\ulm}{\underline{\gl}}
\nc{\Lip}{\on{Lip}}
\nc{\Was}{\on{Was}}
\nc{\tb}{\tilde{b}}

\begin{document}
\title[Damage segregation]
{Damage segregation at fissioning may increase growth rates: A superprocess model}
\author{Steven N. Evans}
\address{Steven N. Evans\\Department of Statistics \#3860\\367 Evans Hall\\University of
  California\\Berkeley, CA 94720-3860\\USA}
\email{evans@stat.berkeley.edu}
\author{David Steinsaltz}
\address{David Steinsaltz\\Department of
  Mathematics and Statistics\\Queen's University\\Kingston, ON K7L 3N6\\CANADA}
  \email{steinsaltz@mast.queensu.ca}
\thanks{SNE supported in part by Grant
DMS-04-05778 from the National Science Foundation.
DS supported in part by Grant K12-AG00981 from the National
Institute on Aging and by a Discovery Grant from the National Science and Engineering Research Council.  }

\subjclass[2000]{}
\date{\today}
\begin{abstract}
A fissioning organism may purge unrepairable damage by bequeathing it preferentially to one of its daughters.  Using
the mathematical formalism of superprocesses, we propose a 
flexible class of analytically tractable models that allow quite general effects of damage on death rates
and splitting rates and similarly general damage segregation mechanisms.  We show 
that, in a suitable regime, the effects of randomness in damage segregation at fissioning 
are indistinguishable from those of randomness in the
mechanism of damage accumulation  during the organism's lifetime.  Moreover, the optimal
population growth is achieved for a particular finite, non-zero level of combined randomness from
these two sources. In particular,
when damage accumulates deterministically, optimal population growth is achieved by a moderately
unequal division of damage between the daughters, while too little or too much division is sub-optimal. Connections are drawn both to recent experimental results on inheritance of damage in unicellular organisms, and to theories of aging and resource division between siblings.
\end{abstract}

\maketitle

\renewcommand{\thefootnote}{*}
\section{Introduction} 
\label{sec:intro}
One of the great challenges in biology is to understand the forces shaping age-related functional decline, termed {\em senescence}.
Much current thinking on senescence ({\em cf}. \cite{repairreplace}) interprets the aging process as an accumulation of organismal damage. The available damage repair mechanisms fall short, it is often argued, because of limitations imposed by natural selection, which may favor early reproduction, even at the cost of later decrepitude. One line of research aims to clarify these trade-offs by examining the non-aging exceptions that test the senescence rule. Of late, it has even been argued that 
negligible \cite{cF90} or even negative \cite{jV04} senescence may not be as theoretically implausible as some had supposed, and that it might not even be terribly rare \cite{jG04}. 

Fissioning unicellular organisms have been generally viewed as a large class of exceptions to the senescence rule. Indeed, their immortality has been considered almost tautological by the principle enunciated by P. Medawar \cite{pM57}, that individual birth is a fundamental prerequisite for aging. This principle has been sharpened by L. Partridge and N. Barton \cite{PB93}, who remark 
\begin{quote}
The critical requirement for the evolution of ageing is that there be a distinction between a parent individual and the smaller offspring for which it provides. If the organism breeds by dividing equally into identical offspring, then the distinction between parent and offspring disappears, the intensity of selection on survival and reproduction will remain constant and individual ageing is not expected to evolve.
\end{quote}
Recent experiments \cite{AGRN03,LJBJ02,caulobacter,SMPT05} have focused attention on the elusive quality of the ``distinction'' between parent and offspring. If aging is the accumulation of unrepaired damage, then ``age'' may go up or down.  The metazoan reproduction that results in one or more young (pristine) offspring, entirely distinct from the old (damaged) parent, is an extreme form of rejuvenation. This may be seen as one end of a continuum of damage segregation mechanisms that include the biased retention of carbonylated proteins in the mother cell of budding yeast \cite{AGRN03} and perhaps the use of aging poles inherited by one of the pair of {\em Escherichia coli} daughter cells as, in the words of C. Stephens \cite{cS05}, cellular ``garbage dumps''. Even where there is no conspicuous morphological distinction between a mother and offspring, the individuals present at the end of a bout of reproduction may not be identical in age, when age is measured in accumulated damage. Whereas traditional theory has focused on the extreme case of an aging parent producing pristine offspring, it now becomes necessary to grapple with the natural-selection implications of strategies along the continuum of damage-sharing between the products of reproduction.

Our approach is a mathematical model of damage-accumulation during a cell's lifetime and damage-segregation 
at reproduction that quantifies (in an idealized context) the costs and benefits of unequal damage allocation to the daughter cells in a fissioning organism. The benefits arise from what G. Bell \cite{gB88} has termed ``exogenous repair'': elimination of damage through lower reproductive success of individuals with higher damage levels. 

For a conceptually simple class of models of population growth that flexibly incorporate quite general structures of damage accumulation, repair and
segregation, we analytically derive the conditions under which increasing inequality in damage inheritance will boost the long-term population growth rate. In particular, for organisms whose lifetime damage accumulation rate is deterministic and positive, some non-zero inequality will always be preferred.  While most immediately relevant for unicellular organisms, this principle and our model may have implications more generally for theories of intergenerational effects, such as transfers of resources and status.

One consequence of exogenous repair may seem surprising: If inherited damage significantly determines the population growth rate, and if damage is split unevenly among the offspring, there may be a positive benefit to accelerating the turnover of generations. In simple branching population growth models, the stable population growth rate is determined solely by the net birth rate. In the model with damage, increasing birth and death rates equally may actually boost the population growth rate. This may be seen as the fission analogue of Hamilton's principle \cite{wH66} linking the likelihood of survival to a given age with future mortality-rate increases and vitality decline, and placing a selective premium on early reproduction. Of course, if the variance in damage accumulation is above the optimum, then this principle implies that reducing the inequality in inheritance, or decreasing birth and death rates equally, would be favored by natural selection.

\section{Background}
Popular reliability models of aging (such as \cite{vK81,GG01,sD82} and additional references in section 3 of \cite{repairreplace}) tend to ignore repair, while the class of growth-reproduction-repair models (such as \cite{soma,AL95,mC97,mM01,CL06} and further references in section 2 of \cite{repairreplace}) tend to ignore the fundamental non-energetic constraints on repair. A living system will inevitably accumulate damage.  Damage-repair mechanisms are available, but these can only slow the process, not prevent it.  The repair mechanism of last resort is selective death.  Any individual line will almost certainly go extinct, but the logic of exponential profusion means that there may still some lines surviving.\footnote{While this argument is simplest for haploid fissioning organisms, the same logic could be applied to the germ line of higher organisms, as in \cite{BP98}.  Sexual recombination, in this picture, only facilitates the rapid diffusion of high-quality genetic material, and the bodies are rebuilt from good raw materials in each generation.}  Following this logic one step further brings us to our central question: What benefit, if any, would accrue to a line of organisms that could not purge damage, but could selectively segregate it into one of its children?

This mechanism of selective segregation of damage
was described by G. Bell \cite{gB88}, who called it ``exogenous repair''.  Bell's theoretical insight derived from his analysis of a class of experiments that were popular a century ago, but have since largely been forgotten.  The general protocol allowed protozoans to grow for a time, until a small sample was plated to fresh medium, and this was repeated for many rounds.  Effectively, this created an artificial selection of a few surviving lines, selected not for maximum fitness (as would be the case if the population were allowed to grow unmolested), but at random.  In most settings, in the absence of sexual recombination, {\em population senescence} --- slackening and eventual cessation of growth --- was the rule. While Bell attributed this decline to M\"uller's ratchet and the accumulation of genetic damage, the same principle could be applied equally to irreparable somatic damage.

One would like to follow the growth of individual cells, but this was inconceivable with the technology of the time.  Only lately has such an undertaking become not only conceivable, but practicable.  Recent work \cite{SMPT05} has followed individual {\em E. coli} over many generations, following the fates of the ``old pole'' cell (the one that has inherited an end that has not been regenerated) and the ``new pole'' cell.  Whereas individual cells do not have a clearly defined age, it makes more sense to ascribe an age to a pole.  Unpublished work   from the same laboratory (described in October 2004 at a workshop held at the Max Planck Institute for Demographic Research in Rostock), as well as the work of \cite{AGRN03,LJBJ02} on {\it Saccharomyces cerevisiae} goes even further, tracking the movement of damaged proteins or mitochondria through the generations, and showing that the growth of the population is maintained by a subpopulation of relatively pristine individuals. The demography of damage accumulation in the fissioning yeast {\it Schizosaccharomyces pombe} has recently been described in \cite{mpiyeast}. These striking experiments have helped focus attention on the elusive nature of the asymmetries that underly microbial aging.

Suppose the ``costs'' of accumulating damage are inevitable, and that only the equality of apportionment between the two daughters may be under evolutionary control.  Of course, perfectly equal splitting of damage is physically impossible, and one must always take care not to read an evolutionary cause into a phenomenon that is inevitable.  At the same time, it is worth posing the question, whether there is an optimal level of asymmetry in the segregation of damage which is non-zero.  If this is true in the models, it suggests possibilities for future experiments, to determine whether the asymmetry is being actively driven by the cell, or whether possibilities for reducing the asymmetry are being neglected. 

In principle, this is a perfectly generic evolutionary phenomenon.  It depends upon the ``old-pole'' experiments only to the extent that they reveal at least one mechanism producing such an asymmetry.  Our model is intended to represent a large population of fissioning organisms, each of which has an individual ``damage level'', which determines its rate of growth and division, and its likelihood of dying.  When a cell divides, its daughters divide the parent's damage unequally, with the inequality controlled by a tunable parameter.  The evolution of this population is described formally by a mathematical object called a ``superprocess''.  (A somewhat different mathematical approach to the growth of structured populations was presented by \cite{DGMT98}. One attempt to model the costs -- but not the possible benefits -- of asymmetries like those found in the Stewart et al. experiments is \cite{JM06}.)

\begin{figure}[!h]
\includegraphics[height=6.5in,width=5in]{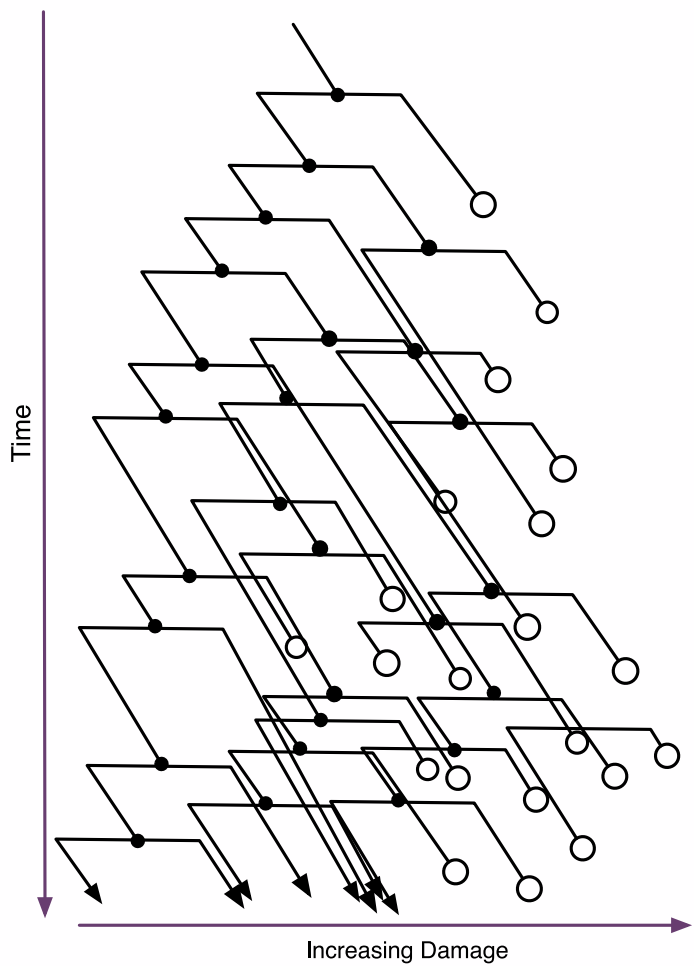}
\caption{Illustration of the branching model.  Sloping lines are lifelines of cells, moving downward in time and rightward in damage space.  Filled circles represent fission events.  Open circles represent deaths.  Arrows represent cells that survive to the end of the experiment.}\label{F:illustration}
\end{figure}

An illustration of the underlying branching model is in Figure \ref{F:illustration}.  In this example, each individual cell accumulates damage at a constant rate during its lifetime, so that the lifelines of cells are parallel sloping lines.  The only way to reduce the damage load is to pass the damage on disproportionately to one of the daughters at a division.  Note the following:
\begin{itemize}
\item Any fixed line of descent eventually dies.
\item The population as a whole continues to grow.
\item Death is more likely, and fissioning comes more slowly, for cells that are further right --- that is, with more damage.
\item The population eventually clusters at low levels of damage.
\end{itemize}  

At this point, we should acknowledge \cite{indianstewart}, which appeared while the present paper was under review. This paper specifically models the old-pole effect, putting forward an account of the possible advantages of asymmetric damage division, which complements our more abstract model in several important respects.
Whereas the present paper offers analytical solutions for models that
incorporate a broad range of mechanisms for damage segregation and the effect of
damage on death and splitting rates, the 
specific parametric model of \cite{indianstewart} is amenable to simulation and computational exploration of its particular 
parameter space. That model is free of the scaling assumptions, on which the analysis given here depends.  (Roughly speaking, the scaling
assumptions are analogous to those that are required for the validity of the diffusion approximations introduced
by researchers such as Kimura into population genetics.) Finally, \cite{indianstewart} differs from the present paper in its explicit representation of cell growth. The important lesson is that the same fundamental behavior is seen in two very different models, analyzed with very different methods.  In each case, increasing the inequality in transmission of damage to the daughter  cells is found to increase the population growth rate in some circumstances.

\nc{\pn}{\P^{(n)}}
\nc{\xnt}{X^{(n)}_{t}}
\nc{\xn}{X^{(n)}}
\section{Description of the model}  \label{sec:branching}
Our model for the growth of a population of fissioning organisms such as {\em E. coli}
is an infinite population measure-valued
diffusion limit of a sequence of finite population branching models.  Mathematical derivations of important properties of this model are left for section \ref{sec:math} and Appendix A. In the present section, we offer a more intuitive description of the model.

\subsection{Diffusion}
The state of the process at any time $t$ is a collection of cells, each of which has some level of damage.  We represent
the level of damage as a positive real number.  Each cell develops independently of its fellows, performing four different behaviors: Damage creation, damage repair, death, and fissioning.  Damage creation and repair sum to a net damage process, which we model as a diffusion.  Diffusion processes, which are most commonly applied in population genetics as models for fluctuating proportions of alleles in a population ({\em cf}. \cite{wE79}), may be thought of as continuous analogues of random walks.  A real-valued diffusion is a general model of a random process that changes continuously in time and satisfies the Markov property: future behavior depends only on the current state, not on the more distant past. The diffusion model of damage accumulation in a cell is determined by two parameters: the diffusion rate $\gs_{\mathrm{mot}}(x)$, giving the intensity of random fluctuation of the damage level as a function of the current damage level $x$; and the drift $b(x)$, giving the trend in damage, whether increasing (positive) or decreasing (negative), as a function of current damage.


\subsection{Branching diffusions and measure-valued diffusions}
While cells move independently through damage space, they are also splitting in two and dying, at rates that depend on their current damage state. The obvious way to represent the state of this ``branching diffusion'' process at any given time is as a list of a changing number of cells, each labeled with its damage state. It turns out to be mathematically more convenient to invert this description 
and present all the damage states, each with a number (possibly zero) representing the number of cells in that state. Formally,
such a description is called an {\em integer-valued measure} on the space of damage states
(that is, on the positive real numbers).   A stochastic process whose  state at any given time is one of these measures is a {\em measure-valued (stochastic) process}.

We take this mathematical simplification one step further, at the expense of increasing the amount of hidden mathematical machinery. The discrete numbers of individuals in the population prevent us from applying the powerful mathematical tools of analysis. This is exactly analogous to the difficulties that arise in analyzing the long-term behavior of discrete branching models, or inheritance models such as the Wright-Fisher and Moran models \cite{wE79}. The famous solution to that problem was W. Feller's \cite{wF51} diffusion approximation. By an essentially analogous method (see, for example, \cite{aE00} or \cite{MR2003k:60104}), rescaling the time and giving each cell ``weight'' $1/n$, while letting the population size and the birth and death rates grow like $n$, and now sending $n$ to infinity, we obtain in the limit 
a stochastic process that has as its state space continuous distributions of ``population'' over the damage states.  Such
general distributions  are called {\em measures} and may be identified 
in our setting with density functions on the positive real numbers.  We stress that the limiting stochastic process
is not deterministic and can be thought of as an inhomogeneous cloud of mass spread over the
positive real numbers that evolves randomly and continuously in time.  Such a stochastic process is called
a {\em measure-valued diffusion} or {\em superprocess}.     An accessible introduction to measure-valued process
produced by this sort of scaling limit, including the original superprocess, called ``super-Brownian motion,'' is \cite{gS02}. 

As described in \cite{sS76}, branching diffusions \cite{pM62,CK70,wE69} and the related stepping-stone models \cite{mK53,KW64} were well established in the 1960s as models for the geographic dispersion, mutation, and selection of rare alleles. Important early applications of spatially structured branching models include \cite{jK77,SF81}, which helped to elucidate the spatial population distributions that are likely to arise from from populations under local dispersion and local control. The utility of measure-valued stochastic processes in population biology was established by the celebrated limit version of allele frequency evolution in a sampling-replacement model due to W. Fleming and M. Viot \cite{flemingviot}, which followed quickly upon the mathematical innovations of D. Dawson \cite{dD75}. In particular, studies such as \cite{EG87,GT99} show how the measured-valued diffusion approach helps in taming the complexities of infinite sites models.  Superprocesses have found broad application in the study of persistence and extinction properties of populations evolving in physical space (such as \cite{wY01,aE04}) or in metaphorical spaces of genotypes (such as \cite{DK99,WB02,DH83}). A thorough introduction to the subject, including extensive notes on its roots and applications in branching-diffusion and sampling-replacement models, is \cite{dD93}.

\subsection{The damage-state superprocess}
Formally, we posit a sequence of branching models, converging to our final superprocess model.  If the scaling assumptions are reasonable, important characteristics of the branching models --- in particular, the connection between the process parameters and the long-term growth rate --- should be well approximated by the corresponding characteristics of the superprocess.

The $n^{\mathrm{th}}$ model in the sequence begins with a possibly random number $N_n$ of individuals.  Individuals are located at points of
the positive real line $\R^+$.  The location of an individual is a measure of the individual's degree of cellular damage.   We code the ensemble of locations
at time $t \ge 0$ as a measure $X^n_t$ on $\R^+$  by placing mass $1/n$ at each location, so that  the initial disposition of masses 
$X^{n}_0$ is a random measure with total mass $N_n/n$.  During its life span the damage level fluctuates, as the organism accumulates and repairs damage.
An individual's level of damage evolves as an independent 
diffusion on $\R^+$ with continuous spatially varying
drift $b(x)$, continuous diffusion rate $\gs_{\mathrm{mot}}(x)$, and continuous killing rate (depending on $n$) $k^{n}(x)$.  While our mathematical framework allows general behavior (mixtures of reflection and killing) at the boundary point 0, it seems sensible for applications to assume complete reflection.  Other boundary behavior would imply a singular mortality mechanism at damage level 0.

An individual can split into two descendants, which happens at spatially varying rate $a^{n}(x)$.   As in the standard Dawson-Watanabe superprocess
construction \cite{EK86,MR94m:60101,aE00}, we assume that birth and death rates grow with $n$ but  are nearly balanced, so that there is an asymptotically finite net birth rate
\begin{equation}  \label{E:netbirth}
\beta(x):=\lnf \lp a^{n}(x)-k^{n}(x)\rp < \infty
\end{equation}
while the total rate at at which either death or splitting occurs satisfies
\begin{equation}  \label{E:branchvar}
\rho(x):=\lnf n^{-1}\left(a^{n}(x)+k^{n}(x)\right) < \infty.
\end{equation}
In other words, an individual at location $x$ attempts to split at a rate which is asymptotically $n \rho(x)$, and such
an attempt is successful (resp. unsuccessful) with a probability that is asymptotically $\frac{1}{2}(1 + \beta(x)/n)$
(resp. $\frac{1}{2}(1 - \beta(x)/n)$).  We assume that  $\beta^{*}:=\sup_x\beta(x)<\infty$ and $\beta_\infty:=\lim_{x\to\infty}\beta(x)$ exists (though it may be $-\infty$).  Both of these conditions hold under the biologically reasonable assumption that $\beta$ is non-increasing
(that is, asymptotic net birth rate is non-increasing in the amount
of damage).

The unbalanced transmission
of damaged material leads to one important difference from the standard Dawson-Watanabe superprocess
construction: When a split occurs the 
two descendants are not born at the same point as the parent.  A parent at location $x$ has descendants at locations $x-Y$ and $x+Y$, where $Y$ is randomly chosen from a distribution $\pi^{n}_{x}$
on $\R^+$.  
We suppose that
\begin{equation}
\label{E:segvar}
\gs^{2}_{\mathrm{seg}}(x) :=  \lnf n \int_{-\infty}^{\infty} y^2 \, d\pi^{n}_{x}(y) < \infty
\end{equation}
for some continuous function $\gs^{2}_{\mathrm{seg}}$.  The quantity $\gs^2_{\mathrm{seg}}(x)$
is a measure of the amount of damage segregation that takes place when a parent with level of damage $x$
splits.

We show in the Appendix that the process $X^n$ converges to a limit process $X$,  as $n \to \infty$, as long as the initial damage distributions $X^n_0$ converge to some finite measure $X_0$. The limit is a Dawson-Watanabe superprocess whose underlying spatial motion is a diffusion with drift $b$ (the same as the original motion) and diffusion rate $\gs$, where
\begin{equation}  \label{E:sigma}
\gs^{2}(x):=\gs^{2}_{\mathrm{mot}}(x)+\gs^{2}_{\mathrm{seg}}(x)\rho(x)
\end{equation} 
Thus the  asymptotic  effect of segregation is equivalent to an increase in the diffusion rate, which may be thought of as the extent of random fluctuation in the the underlying motion of an individual cell through damage space. The same effect may be achieved by increasing simultaneously the fission and death rates, leaving fixed the net rate of birth.  The quantity $\gs^{2}_{\mathrm{seg}}(x)$
measures a re-scaled per generation degree of damage segregation, and multiplication by the re-scaled splitting rate $\rho(x)$
produces a quantity $\gs^{2}_{\mathrm{seg}}(x)\rho(x)$ that measures a re-scaled per unit time rate of damage segregation.

\section{Major results}
Write $Z_t$ for the asymptotic re-scaled population size at time $t$.  That is, $Z_t$ is the total mass of the measure
$X_t$.  We show that there is an {\em asymptotic population growth rate} $\ulm$ in the sense
that the asymptotic behavior of the expectation of $Z_t$ satisfies
\begin{equation}  \label{E:agr}
\ltf t^{-1}\log \E \lb Z_t \rb=\ulm.
\end{equation}
Thus, $\E \lb Z_t \rb$ grows to first order like $e^{\ulm t}$.

When the asymptotic growth rate $\ulm$ is positive, this result may be significantly strengthened.  
Theorem \ref{T:rescaling} tells us then (under mild technical conditions) that:
\begin{itemize}
\item To first order, the total population size at time $t$, $Z_t$, grows like $e^{\ulm t}$.  More precisely, 
$Z_t / \E \lb Z_t \rb$ converges to a finite, non-zero, random limit as $t \rightarrow \infty$.
\item The random distribution describing the relative proportions of damage states of cells at time $t$ converges to 
a non-random distribution with density $c \phi$ as $t \rightarrow \infty$, 
where $\phi$ is the unique solution to the ordinary differential equation
\begin{align*}
\oh\left[\gs^{2}(x)\phi(x)\right]'' -&\left[b(x) \phi(x)\right]' +\beta(x)\phi(x) =\gl \phi\\
 &\text{with } 
\frac{\gs(0)^{2}}{2} \phi'(0)= 2b(0)\phi(0)
\end{align*} 
and $c$ is the normalization constant required to make $c \phi$ integrate $1$.
That is, for any $0 \le a < b < \infty$ the proportion
of the population with damage states between $a$ and $b$  converges
to $\int_a^b \phi(x) \, dx / \int_0^\infty \phi(x) \, dx$.
\end{itemize}

Suppose now that we identify (as is commonly done) fitness of a cell type with the asymptotic growth rate $\ulm$.
It is not obvious, under arbitrary fixed conditions of average damage accumulation and repair
($b$), splitting rate ($\rho$), and net reproduction ($\beta$), what level of randomness in damage accumulation
and repair ($\sigma_{\mathrm{mot}}$) and damage segregation ($\sigma_{\mathrm{seg}}$) has the highest fitness.  Under the biologically reasonable
assumption that the average damage accumulation rate is everywhere positive, though, it is generally true that moderate degrees of randomness in both mechanisms lead to
higher growth rates than either very small or very large degrees
of randomness.  Theorem \ref{T:optimal} (in section \ref{sec:highdif}) tells us that, under fairly general conditions, if the diffusion rate $\gs$ is reduced uniformly to 0, or increased uniformly to $\infty$, this pushes the asymptotic growth rate down to $\beta_{\infty}=\lim_{z\to\infty}\beta(z)$.

Since the growth rate at any damage level is higher than $\beta_\infty$, this result tells us that the growth rate is minimized by sending $\gs$ uniformly either to 0 or to $\infty$.  Too much or too little randomness in damage accumulation yields lower growth rates; optimal growth must be found at intermediate levels of diffusion.  

Imagine now a situation in which the upward drift in damage has been brought as low as possible by natural selection, optimizing repair processes and tamping down damage generation.  We see that there may still be some latitude for natural selection to increase the lineage's growth rate, by tinkering with this randomness. Recall that $\gs$ is composed of two parts: the randomness of damage accumulation during the lifecourse, and the inequality of damage transmission to the daughters.  Without altering damage generation or repair, but only the equality of transmission to the offspring, the cell type's long-term growth rate may be increased.  In particular, if damage accumulation rates are positive and deterministic --- a fixed rate $b(x)$ of damage accumulation per unit of time, when a cell has damage level $x$ --- the growth rate will always be improved by unequal damage transmission.

\section{An example}  \label{sec:example}
Suppose that damage accumulates deterministically during the life of an organism, at a rate proportional to the current level of damage.  In the absence of damage segregation, then, the damage at time $t$ grows like $e^{bt}$ (this corresponds
to $b(x) = bx$ and $\sigma^2_{\mathrm{mot}} = 0$).  We suppose that the combined parameter $\sigma^2_{\mathrm{seg}} \rho(x)$
describing the time rate of damage segregation for an organism that fissions at damage level $x$ is $\tau^2 x^2$, where $\tau$ is a tunable non-negative parameter (hence, 
$\sigma(x) = \sqrt{\sigma^2_{\mathrm{seg}} \rho(x)} = \tau x$).  The
generator for the spatial motion in the limit superprocess is thus
\[
\frac{\tau^2}{2} x^2 f''(x) + b x f'(x).
\]
This is the generator of the geometric Brownian motion 
$\exp(\tau B_t + (b - \frac{\tau^2}{2}) t)$.  We see immediately that adding enough noise can push the spatial motion toward regions with less damage.
To make this fit formally into our theorems, we assume complete reflection at 1. (We have previously taken the range of damage
states to be the positive real numbers $(0,\infty)$, whereas here we have taken the interval $(1,\infty)$. 
The choice of interval is irrelevant as far as the mathematics goes,
 and we could convert this choice to our previous choice by simply re-defining $b(x)$ and $\sigma(x)$).

Suppose that the bias of reproduction over dying in the superprocess is of the form $\beta(x)=\beta_0 - \beta_1 x$.  The population growth rate will be the same as for the process $Y_{t}=B_{t}+(\frac{b}{\tau}-\frac{\tau}{2})t$, with reflection at 0, and killing at rate $\beta_{0}-\beta_{1}e^{\tau Y_{t}}$.  The eigenvalue equation is
\[
\frac{1}{2} f''(y) + \tb x f'(y) + (\beta_0 - \beta_1 e^{\tau y}) f(y) = \lambda f(y),
\]
where $\tb:=\frac{b}{\tau}-\frac{\tau}{2}$, with boundary condition $f'(0)=0$.

Using standard transformations as in section 6 of \cite{quasistat}, or symbolic mathematics software such as {\em Maple}, we may compute  a basis for the space of solutions to be the modified Bessel functions
\[
I_{\nu}\left(\sqrt{8\beta_1/\tau^{2}} e^{\tau y/2}\right)  e^{-\tb y} \text{ and }
K_{\nu}\left(\sqrt{8\beta_1/\tau^{2}} e^{\tau y/2}\right)  e^{-\tb y},
\]
where
\begin{equation}  \label{E:nu}
\nu=\frac{2\sqrt{\tb^{2}-2(\beta_{0}-\gl)}}{\tau}.
\end{equation}

\begin{figure}[h]
\includegraphics[height=5in,width=5.5in]{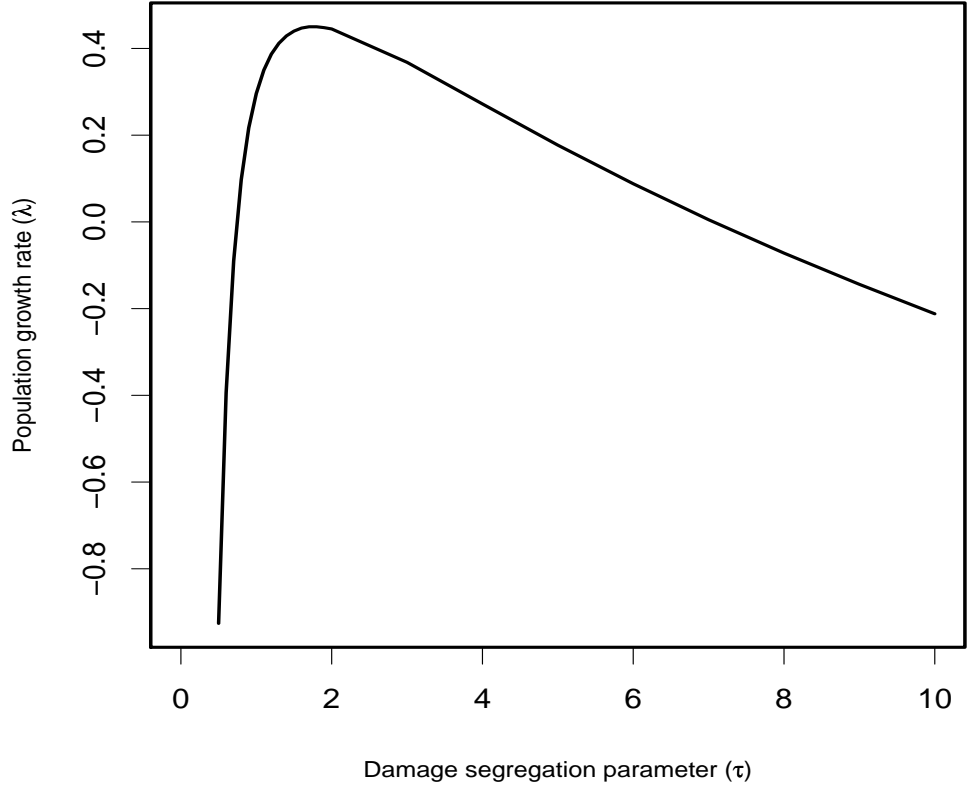}
\caption{Asymptotic growth rates $\ulm$ computed from \eqref{E:ulm} for populations whose deterministic rate of damage accumulation is equal to the current damage level, and whose relative excess of births over deaths is $1.9-x/2$ at damage level $x$.  Damage segregation is added at rate $\tau x$ for births to parents at damage level $x$.}\label{F:example}
\end{figure}

\begin{figure}[h]
\includegraphics[height=5in,width=5.5in]{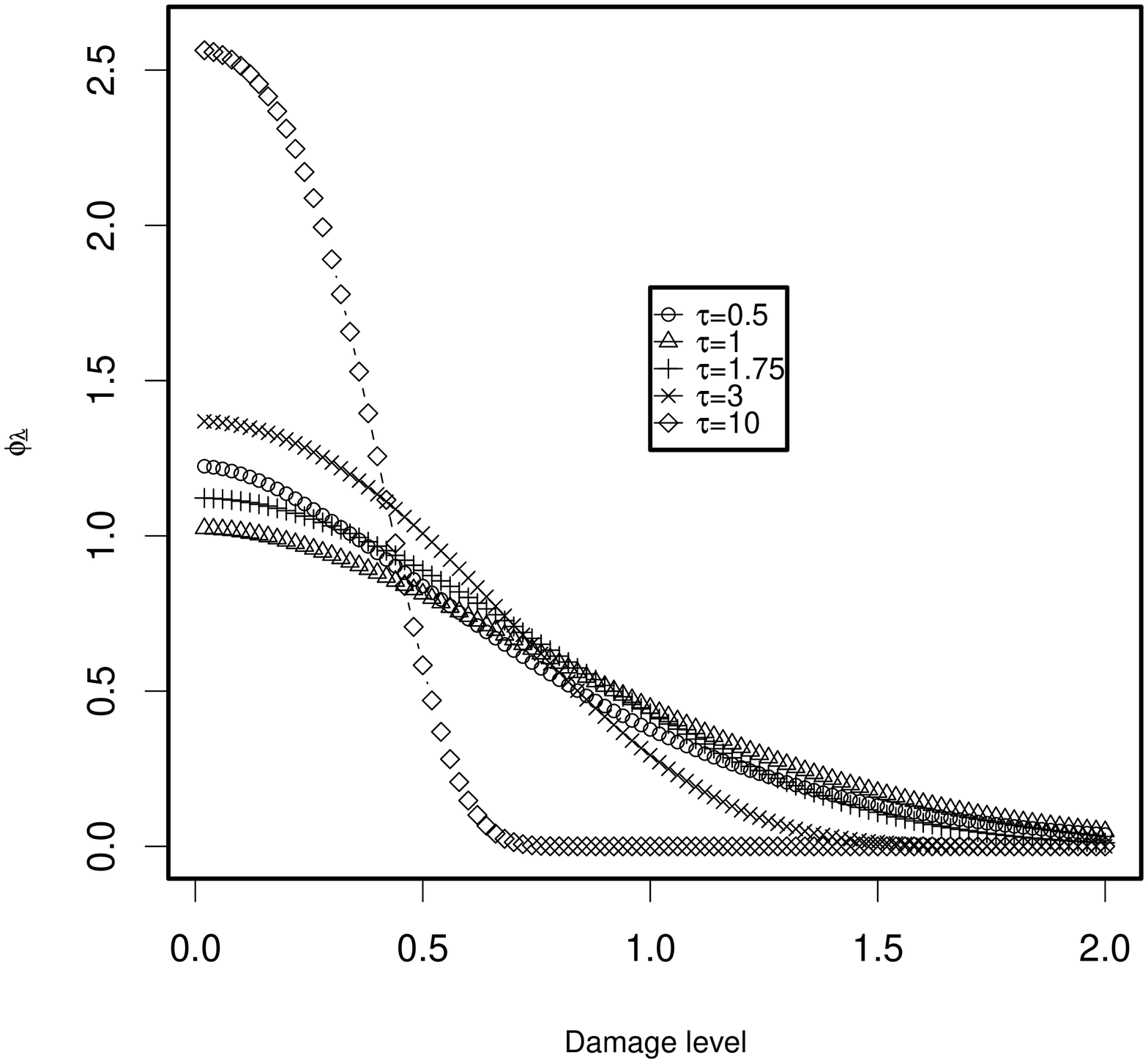}
\caption{Asymptotic distribution of damage for the choice of parameters $\beta_{1}=\oh$, $b=1$ and $\beta_{0}=1.9$, and various choices of $\tau$.}\label{F:example2}
\end{figure}

By definition, $\ulm$ is the infimum of those $\gl$ such that the solution $f$ is non-negative.  Since $f'(0)=0$, it cannot be the case that $f(0)=0$.  Hence, the solution for $\ulm$ has no zero, but the solution for every $\gl<\ulm$ does have a zero in $\R^{+}$.  By continuity (as argued for the same problem in \cite{quasistat}), it follows that the solution for $\ulm$ is a multiple of $K_{\nu}$.
The boundary condition translates to 
\begin{equation}  \label{E:Knu}
\frac{K'_{\nu}}{K_{\nu}}\left(\frac{\sqrt{8\beta_{1}}}{\tau}\right)=\frac{\tb}{\sqrt{2\kappa_{1}}}.
\end{equation}

\nc{\un}{\underline{\nu}}

Find $\nu_{0}$ real such that the last zero of $K_{i\nu_{0}}$ is at $\sqrt{8\beta_{1}}/\tau$, and represent the corresponding $\gl$ by
$$
\gl_{0}=-\frac{\tau^{2}\nu_{0}^{2}}{8}-\frac{\tb^{2}}{2}+\beta_{0}.
$$
Then
$$
\ulm=\inf\left\{ \gl\in (\gl_{0},\infty): \frac{K'_{\nu}}{K_{\nu}}\left(\frac{\sqrt{8\beta_{1}}}{\tau}\right)=-\frac{\tb}{\sqrt{2\beta_{1}}}\right\}.
$$

As an example, take $\beta_{1}=\oh$, $b=1$ and $\beta_{0}=1.9$.  Using the mathematical programming language {\em Maple} we compute the asymptotic growth rates given in Figure \ref{F:example}.  The maximum growth rate of occurs at $\tau=1.75$. Figure \ref{F:example2} shows the asymptotic distribution of damage for different values of $\tau$.  Note that as $\tau$ increases, the asymptotic distribution becomes increasingly compressed into the lower range of damages.  

Remember that $\gs=\sqrt{\gs^{2}_{\mathrm{mot}}+\gs^{2}_{\mathrm{seg}}\rho}$, where $\gs^{2}_{\mathrm{mot}}$ quantifies the randomness of fluctuations in the damage during the life of a cell; $\gs^{2}_{\mathrm{seg}}$ quantifies the extent to which the damage is split unequally between the daughters and $\rho$ is the splitting rate. More unequal distribution of damage among the daughter cells means that there will be effectively more randomness in the rate of damage accumulation within a line.

We point out one implication for the evolution of generation time.  Suppose that there is no mechanism available for altering the extent of damage segregation in each fission event.  (Perhaps damage segregation is not actively determined, but is only a consequence of the inevitable one-way inheritance of an ``old pole'', or simple random fluctuation.)  Suppose, though, that the organism can accelerate the generations. This has the effect of increasing $\gs$, which may be enough to shift the asymptotic growth rate into the positive. 

To be specific, suppose there is a population of $1000$ bacteria, with deterministic damage accumulation at a rate $x$ units per day when the cell has $x$ units of damage.  A cell with $x$ units of damage has a hazard rate for reproduction in the next instant of $25.95-x/4$, and a hazard rate for dying of $24.05+x/4$. (Measuring time in days, this corresponds to a mean time between fission events of about 1 hour.) In the notation of our model, this means that $\rho(x)=50/1000=0.05$ and $\beta(x)=1.9-x/2$.  The minimum level of damage in a cell is 1 unit. At a fission event, if the mother has $x$ units of damage, one daughter gets 90\% of $x$ and the other 110\% of $x$. (Damage units must be thought of as proportional to the total volume of the cell. Thus, when a cell with $x$ damage units divides, there is $2x$ damage to be divided up.) From equation \eqref{E:segvar}, we compute that $\gs_{\mathrm{seg}}(x)=3.1x$. Since $\gs_{\mathrm{mot}}$ was assumed to be 0, this means that $\gs(x)=\gs_{\mathrm{seg}}(x)\sqrt{.05}=0.7x$. We see from Figure \ref{F:example} that the asymptotic growth rate of the population will be $-0.09$, meaning that the population may be expected to decline over time. 

Suppose now there were a mutant strain which had boosted its rate of reproduction to $40-x/4$, but was penalized with a rate of dying $38.2+x/4$. That is, fission rates were increased by 14.05 (about 35\% at the minimum damage state), but the mortality rate was increased by a bit more, by 14.15. Everything else stays the same. We now have $\rho=78.2/1000=.0782$, meaning that $\gs(x)=0.87x$, and despite the mortality penalty we see that the population now has a positive asymptotic growth rate of 0.09. Instead of declining, this strain will tend to double in numbers about every 10 days.

\section{Conclusions}
We have introduced a mathematical model of population growth, for a population of haploid organisms that accumulate damage.  This is a limit of branching processes, in which individuals accumulate damage, rising and falling at random, with an upward tendency.  Under fairly general assumptions we have shown
\begin{itemize}
\item The population growth rate converges to a fixed rate, determined by the solutions to an ordinary differential equation (ODE);
\item If the asymptotic growth rate is positive, the relative proportions of damage levels within the population converge to a fixed distribution, also given by the solution to an ODE;
\item The effect of increasing damage segregation in the model (that is, causing one daughter cell to have higher damage than the parent after the split and one lower) is equivalent to increasing the randomness (diffusion) of the damage-accumulation process;
\item Accelerating the turnover of generations --- increasing birth and death rates equally, leaving the net birth rate unchanged --- effectively increases damage segregation;
\item In general, the optimum level of combined damage diffusion and damage segregation (as measured by the growth rate) is not 0, but is some finite non-zero level.
\end{itemize}

This is one possible mathematical analysis of exogenous repair.  A strain of cells with deterministic damage accumulation will eventually run itself to extinction because of the increasing damage load.  Perhaps surprisingly, the simple expedient of dividing the damage unequally between the offspring may be enough to rescue the line.  In general, if the damage diffusion rate were sub-optimal, a mutant line could obtain a selective advantage by increasing the level of damage segregation.  In fact, damage segregation could be a worthwhile investment even if it came at the cost of a slight overall increase in damage accumulation. Perhaps even more surprising, simply accelerating cell division and cell death equally, leaving the net birth rate unchanged, may be enough to shift a negative population growth rate into the positive range. Of course, there can also be too much damage segregation, and there is potential for improving growth rates by equalizing the inheritance between the daughter cells, if the variation in damage accumulation is excessive.

An abstract model such as presented in this paper is primarily a guide to thinking about experimental results, rather than a template for analyzing the data from any given experiment.  This does not mean that this model cannot be experimentally tested.  Some general predictions of the model are:
\begin{enumerate}
\item Mechanisms for actively increasing damage segregation in reproduction should be common.
\item While individual lines tend toward increasing damage, the overall distribution of damage in the population stabilizes.
\item Damage segregation compensates for ``too little'' randomness in damage accumulation. Thus more unequal division of damage might be expected in organisms whose rate of damage accumulation is more deterministic.
\item Manipulating the extent of damage segregation should affect the population growth rate.  Under some circumstances, increasing segregation should increase the growth rate.
\item One way of manipulating damage segregation is to manipulate the generation time. Increasing the fission rate and the death rate equally has the effect of increasing the segregation effect.
\end{enumerate}

The first ``prediction'' is already fairly established. While asymmetric reproduction and exogenous repair were implicit in the century-old line of experiments recounted in \cite{gB88}, the direct measure of asymmetry in unicellular reproduction has only recently become possible. Some of the organisms studied are
\begin{enumerate}
\item {\em Caulobacter cresentus}: This bacterium grows from one end and then divides. The old end accumulates damage and senesces, while the new end experiences ``rejuvenating reproduction'' \cite{caulobacter}.
\item {\em Escherichia coli}: While this bacterium seems to divide symmetrically, rejuvenation seems to happen preferentially in the middle, while the poles accumulate damage. \cite{SMPT05} found declining vigor with increasing age of the inherited cellular poles.
\item {\em Saccharomyces cerevisiae}: The aging of the mother cell in these budding yeast was established as long ago as \cite{MJ59}, and the functional asymmetry between mother and daughter was established by \cite{HU77}. \cite{LJBJ02} showed that the continuing lineage growth depends upon maintaining the asymmetry in damage state between mother (more damaged) and daughter (less damaged).  An {\em atp2} mutant strain that failed to segregate active mitochondria preferentially to the daughter cell succumbed to clonal senescence.  \cite{AGRN03} found that carbonylated proteins are selectively retained by the mother cell, and that this requires an active SIR2 gene.  (Higher activity of SIR2 also increases the longevity of the mother cell itself, which seems surprising on the face of it; but this is likely a consequence of epistatic effects, {\em cf}. \cite{KMG99}.)
\item {\em Schizosaccharomyces pombe}: These fissioning yeast seem superficially to divide symmetrically. While there is no gross morphological distinction on the basis of which the two products of reproduction could be allocated to categories ``parent'' and ``offspring,'' the fission, there is heritable asymmetry between the two fission products in size \cite{BW99} and number of fission scars \cite{CZJY80}. The inheritance of damaged proteins was studied in \cite{mpiyeast}, but asymmetry of inheritance was not measured directly.
\end{enumerate}

Yeast seem to be a promising class of organisms for studying the inheritance of damage. As remarked by Lai {\it et al.} \cite{LJBJ02},
\begin{quote}
The
aging process of yeast cells best illustrates the cellular
and generational asymmetry [\dots.] The immortality of the yeast clone] depends upon the establishment of an age difference between a daughter cell and the mother cell that produces it. Although 
the daughter cell receives cellular components from the 
mother cell, it does not inherit the mother cellÕs age and is always younger than the mother cell. This age 
asymmetry implies that any substance or process responsible for the aging of mother cells must be carefully isolated from the daughter cells.
\end{quote}
The biodemography of both {\em S. cerevisiae} and {\em S. pombe} is relatively well understood, and methods for tracking fission scars, carbonylated proteins and inactive mitochondria are well developed. We have two species with quite different types of damage segregation, and comparisons are likely to prove illuminating.
The identification of other clonal senescence mutants that act on damage segregation, whether variant mutations to SIR2 or another gene, would expand the toolkit for tuning damage segregation in {\em S. pombe}.

The recent study of carbonylation \cite{mpiyeast} seems particularly amenable to analysis in terms of exogenous repair and our model. This study examined the distribution of damaged proteins in the whole population, finding that the dividing cells were concentrated down near 0, while the non-dividing cells had a significantly higher level of damage. The main parameters in our model --- the rate of damage accumulation in stationary phase, the vital rates for cells with a given damage level, the type of damage inheritance --- could all be measured in principle. Indeed, the symmetry of damage inheritance between the daughters was examined, but only to point out that there is some sharing of damage, that the carbonylated proteins are not all retained by one of the two daughters. Pushing this further, it would be possible to measure the inequality of damage inheritance, both of carbonylated proteins and inactive mitochondria. Once these parameters had been estimated, it would be possible to compare the damage distribution in the paper with our theoretically predicted asymptotic distribution. 

Comparing mutant strains with different damage segregation properties would further refine the comparison. It would also be possible to perform competition experiments: Broadly speaking, we would predict that as the overall damage load increases --- perhaps by exposure to paraquat or disabling of antioxidants --- this would shift the optimal level of damage segregation.  In some simple choices of parameters for the model this shift would be toward more extreme damage segregation, the same effect that would be predicted from decreasing the variability of damage accumulation.  Thus, we might expect that strains exposed to such toxic environments over some time would evolve to show higher levels of damage segregation. If damage segregation is not controlled by active mechanisms, this means that increasing damage accumulation should select for a faster turnover of generations: simultaneously higher rates of cell death and fissioning.

Perhaps no simple resolution should be expected.  We are examining the interaction of individual-level aging downward, with aging of parts of individuals, and simultaneously upward, toward population-level aging.  In the present model one might say that there is a partial individualization of senescence.  That is, while there is no identifiable individual who ages and dies, leaving behind youthful descendants, damage segregation has transformed the senescence problem plaguing the entire population into a problem of some individual cells.  

There has been some controversy (see \cite{cW06}, and the response \cite{ST05}) about whether results such as those of \cite{SMPT05} reflect genuine ``aging.'' Perhaps more profitable is to see, in this growing family of experimental results, clues to the broader context of aging: Paradigmatic aging in our species and similar ones is one of a class of mechanisms which function to redistribute damage within a population. Most metazoans, perhaps inevitably, have converged on the extreme strategy of keeping essentially all the damage (even generating much more damage in the process), and produce pristine offspring. In unicellular organisms, there seem to be a broader range of mechanisms, and degrees of damage differentiation in the end-products of reproduction. 

Is this relativization of senescence of any relevance for metazoans, then? It is, for at least two separate reasons. The first is that the propagation of the germ line is in many senses comparable to the propagation of a line of unicellular organisms. As eloquently commented by Lai {\it et al.} \cite{LJBJ02},
``Some of the key questions in aging concern the differences between germline and soma. Any mechanism invoked to explain the aging of the soma must also be able to accommodate the immortality of the germline.
At the level of the organism, the issue is the renewal
that occurs at each generation, providing the progeny
a fresh start with the potential for a full life span.'' This principle may also be significant for the regulation of replicative senescence in somatic stem cells or other somatic cells. The recent discovery of non-random segregation in mouse mitotic chromosome replication \cite{ArmKla06,AK07} may be a hint in this direction. Even more suggestive is the apparent functioning of this mechanism in maintaining the integrity of cell lines during the lives of individual higher eukaryotes (small intestine epithelial crypt cells in {\em H. sapiens} and neural precursor cells in {\em D. melanogaster}) \cite{RBS06}.

The other reason is perhaps more profound, and certainly more speculative. Over the past several decades, mathematical theory has played a significant part in the evolving discussion of risk-management and bet-hedging in natural selection.  J. Gillespie's seminal treatment \cite{jG74} notes that
\begin{quote}
the variance in the numbers of offspring of a genotype has two components, the within-generation component resulting from different individuals of the same genotype having different numbers of offspring, and a between-generation component due to the effects of a changing environment.
\end{quote}
The general conclusion has been that lower variance in essential demographic traits is selectively favored \cite{FS90,dG84}, although this picture is complicated by varying environments, if the variance is associated with variable response to the environment \cite{EH94}. Damage accumulation falls into a third class of variance: variable epigenetic inheritance of a resource shared between offspring. 

Our model departs from the models of within-generation variation in which the varying phenotypes of siblings are independent, but is allied with models of resource sharing within families.  In the case of damage-accumulation the ``resource'' is the undamaged cell components, but this model might in principle be applied to the transmission of resources or status from parents to offspring.  In contrast to the general selective advantage for reduced  within-generation variability in the uncorrelated setting, the results of our analysis suggest that increased variability may be selectively favored when the phenotype is a shared resource. If the resource is simply size, then our model might be compared to the asymmetric-division cell cycle model developed by \cite{jT89}. That work lacked an explicit stochastic population model, and did not address the question of overall population growth rate, but some of the observations from that paper may be relevant, most significantly the simple empirical fact of size asymmetry in yeast division.

This beneficial variance may have some relevance for the growing recognition that the evolution of longevity depends fundamentally on the nature of intergenerational transfers.  All the work on this problem of which we are aware, particularly \cite{rL03,KR02,RK03,CL06}, implicitly assumes that resources are divided equally according to need (though \cite{HDS02} points out that the goal of equality may still produce systematic biases based on birth-order).  The key paper of Trivers \cite{Trivers74} explicitly opposes a parental goal of equality against the offspring's interest in monopolizing resources. Models of offspring provisioning in this tradition, such as \cite{PML89,RHP02}, seem to assume a ``diminishing returns'' to investment, hence higher fitness for reduced spread in offspring quality, except when there is a threshold for survival that only one is likely to cross. Our model suggests that the cumulative effect of exponential growth can easily make unequal division of resources a winning strategy. This inequality may be purely random, and need not depend on sibling competition or some other strategy for identifying and rewarding the inherently superior offspring.

We note that intergenerational transfer theory must address the complication of parent-offspring and sibling competition, because of the genetic differences between relatives. Our model avoids those complications, assuming a complete comity of cells.
The parent gives everything to its offspring, and the daughters are as interested in the other's welfare as in their own. Nonetheless --- and this is what may be surprising --- equal endowment of offspring may not be optimal. A daughter's fitness may be best served by trading the sure 50\% inheritance for a lottery ticket that may yield more, although it may yield less.

\section{Mathematical methods} \label{sec:math}
Define $(P_{t})_{t \ge 0}$ to be the semigroup generated by
$$
\mL\phi:=\oh\gs^{2} \phi'' + b \phi' +\beta \phi
$$
with boundary condition
\begin{equation}  \label{E:ivp}
p_{0}\frac{\gs(0)^{2}}{2} \phi'(0)=(1-p_{0}) \phi(0)
\end{equation}
for some fixed $0 \le p_0 \le 1$. (This semigroup will not, in general, 
be sub-Markovian because $\beta$ can take positive values).
As we observe in Section \ref{sec:rescaling}, $\E[\langle \phi, X(t)\rangle] = \int_{\R^+} P_t \phi(x) \, d\nu(x)$ when
$X_0$ is the deterministic measure $\nu$, and so the behavior of $(X_t)_{t \ge 0}$ is governed by
that of $(P_t)_{t \ge 0}$, at least at the level of expectations.  The choice $p_0=1$ is most relevant for
the description of fissioning organisms, because it corresponds to complete reflection at the state
$0$ representing no damage.  Other choices of $p_0$ would imply the rather anomalous presence
of an additional singular killing mechanism at $0$.  However, since the mathematical development
is unchanged by the assumption of a general boundary condition, we do not specialize to
complete reflection.

It turns out that the entire long-term behavior of $(X_t)_{t \ge 0}$ is rather simply connected to that of
$(P_{t})_{t \ge 0}$.  Let
$(P^{*}_{t})_{t \ge 0}$ be the adjoint semigroup  generated by
$$
\mL^{*}\phi(x) :=\oh\frac{d}{dx^{2}}\left[\gs^{2}(x)\phi(x)\right] - \frac{d}{dx}\left[b(x) \phi(x)\right] +\beta(x)\phi(x).
$$
with boundary condition
\begin{equation}  \label{E:ivp2}
p_{0}\frac{\gs(0)^{2}}{2} \lp \phi'(0) - 2b(0)\phi(0)\rp=(1-p_{0})\phi(0).
\end{equation}
For any $\gl$, define $\phi_{\gl}$ to be the unique solution to the
initial value problem
\begin{equation}  \label{E:ivp3}
\mL^* \phi_\gl =\gl \phi_{\gl}\quad
 \text{with }  \phi_{\gl}(0)=p_{0} \text{ and } 
\frac{\gs(0)^{2}}{2}\lp \phi'_{\gl}(0)- 2b(0)\phi_{\gl}(0)\rp=1-p_{0}.
\end{equation} 
Put
\begin{equation}  \label{E:ulm}
\ulm:=\inf\ls \gl \, :\, \phi_{\gl}(x) \text{ is non-negative for all }x \rs.
\end{equation}
We recall in Proposition \ref{P:omegat} that, under suitable hypotheses, for a
compactly supported finite measure $\nu$
\[
\ltf \frac{\langle P_{t+s}\indic,\nu\rangle}{\langle P_{t}\indic,\nu\rangle} = e^{\ulm s}
\]
so that {\em a fortiori}
\[
\ltf t^{-1} \log \langle P_{t}\indic,\nu\rangle = \ulm,
\]
and, moreover,
\[
\ltf \frac{\langle P_{t}\varphi,\nu\rangle}{\langle P_{t}\indic,\nu\rangle} = \frac{\izf \varphi(z)\phi_{\ulm}(z)dz}{\izf \phi_{\ulm}(z)dz}
\]
for a bounded test function $\varphi$.
We subsequently show in Theorem \ref{T:rescaling}, again under suitable hypotheses, that if $\ulm>0$, then
\[
\lim_{t \to \infty} \frac{\langle \varphi, X_t \rangle}{\langle P_{t}\indic,\nu\rangle} = W \frac{\izf \varphi(z)\phi_{\ulm}(z)dz}{\izf \phi_{\ulm}(z)dz}
\]
in probability, where $W$ is a non-trivial random variable (that doesn't depend on $\varphi$).  In particular,
the asymptotic growth rate of the total mass 
$\langle \indic, X_t \rangle$ is $\ulm$ and 
\[
\lim_{t \to \infty} \frac{\langle \varphi, X_t \rangle}{\langle \indic, X_t \rangle} = \frac{\izf \varphi(z)\phi_{\ulm}(z)dz}{\izf \phi_{\ulm}(z)dz}
\]

We show in Section \ref{sec:highdif} under quite
general conditions (including the assumption
that $\beta$ is non-increasing so that
$\beta_\infty = \inf_x \beta(x)$) that the presence of some
randomness in either the damage and accumulation
diffusion or the damage segregation mechanism
is beneficial for the long-term growth of the population
but too much randomness is counterproductive.
That is, if $b$, $\beta$ and $\rho$ are held fixed,
then $\ulm$ converges to $\beta_\infty$ as the diffusion rate function
$\sigma = \sqrt{\sigma_{\mathrm{mot}}^2 + \sigma_{\mathrm{seg}}^2 \rho}$ goes to either $0$ or $\infty$,
whereas $\ulm > \beta_\infty$ for finite $\sigma$.  
%

\subsection{Results from \cite{quasistat}}  \label{sec:quasistat}
Essential to linking the long-term growth behaviour of the superprocess to the theory of quasistationary distributions is the observation that
$\mL-\beta^{*}$ (recall that $\beta^{*}$ is a constant, the upper bound on $\beta$) is the generator of a killed diffusion semigroup $Q_{t}\phi=e^{-\beta^{*}t}P_{t}\phi$.  Thus, when we have a result stating that $Q_{t}$ converges to a stationary distribution with density 
a multiple of $\phi_{\ulm}$, meaning that for any bounded test function $\varphi:\R^{+}$, and any compactly supported finite measure $\nu$,
$$
\ltf \frac{\langle Q_{t}\varphi,\nu\rangle}{\langle Q_{t}\indic,\nu\rangle} = \frac{\izf \varphi(z)\phi_{\ulm}(z)dz}{\izf \phi_{\ulm}(z)dz},
$$
then the same holds for $P_{t}$.

For $t \ge 0$ and $\nu$ a compactly supported finite measure on $\R^+$, set
\begin{equation}  \label{E:omega}
\omega_{t}(\nu):= \frac{\langle P_{t}\indic,\nu\rangle}{P_t \indic(1)}.
\end{equation}  
When $\nu=\delta_{x}$, write $\omega_{t}(x)$.  Note that 
\begin{equation}  \label{E:omeganu}
\omega_{t}(\nu)=\izf \omega_{t}(x) \, d\nu(x)
\end{equation}
for any compactly supported finite measure $\nu$.  Define
\begin{equation}  \label{E:omegastar}
\omega_{*}(x,\nu)=\sup_{t\ge 0} \frac{\omega_{t}(x)}{\omega_{t}(\nu)}.
\end{equation}

Assume the following condition:  For  all compactly supported finite measures $\nu$ on $\R^{+}$, and some $\gep>0$
\begin{equation*}  \tag{GB2}
\sup_{s\ge 0} \sup_{x \in \R^+} e^{-(2\ulm-\gep)s}  \int \omega_{*}(y,\nu)^{2} P_s(x,dy) < \infty.
\end{equation*}

While (GB2) is an abstract condition, it is implied by a fairly concrete condition on the diffusion parameters.

 
\begin{Lem}
\label{L:omegabounded}
Condition (GB2) always holds if the net birth rate  $\beta$ is monotone non-increasing, and there is complete reflection at $0$.
\end{Lem}

\begin{proof}
Changing $\beta$ by a constant leaves $\omega_{t}$, hence $\omega_{*}$ as well, unchanged.  Thus, we may assume that $\beta\le 0$, so that $P_{t}$ is the semigroup of a killed diffusion.

Given $x<x'$, we may couple the diffusion $X_{t}$ started at $x$ with a diffusion $X'_{t}$, with identical dynamics but started at $x'$, so that $X_{t}\le X'_{t}$ for all $t$.  By the monotonicity of $\beta$, this implies that $P_{t}\indic (x)\ge P_{t}\indic(x')$, since $P_{t}\indic(x)=\P_{x}\{\tp>t\}.$  Consequently, $\omega_{t}$ is monotonically non-increasing, and so is $\omega_{*}$.  In particular, for all $y$, $\omega_{*}(y,\nu)\le \omega_{*}(0,\nu)$, which is finite, by the results of section 2.5 of \cite{quasistat}.  Then
$$
\sup_{s\ge 0} \sup_{x \in K} e^{-(2\ulm-\gep)s}  \int \omega_{*}(y,\nu)^{2} P_s(x,dy) \le \omega_{*}(0,\nu)^{2}<\infty.
$$
\end{proof}

We will also need to assume the technical condition
\begin{equation*}
\tag{LP}  
\mL^* \text{ is in the limit-point case at }\infty.
\end{equation*}

We define
$$
\tilde{b}(z):=\frac{b\circ F^{-1}(x)}{\gs\circ F^{-1}(x)} - \oh\gs'\circ F^{-1}(x),
$$
and $\tilde{\beta}(x):=\beta\circ F^{-1}(x)$, where $F(y):=\int_{0}^{y}du/\gs(u)$ is the Liouville transform, the spatial transformation that converts the diffusion coefficient into a constant.  Then Lemma 2.1 of \cite{quasistat} says that (LP) holds whenever
\begin{equation*}
        \liminf_{x\to\infty} x^{-2} \lp \tilde{b}(x)^{2} + \tilde{b}'(x) + 2\tilde{\beta}(x) 
        \rp >-\infty.\tag{LP'}
    \end{equation*}
Examples which do not satisfy this condition must have pathological fluctuations in $b$.

Define $\psi_{\gl}$ to be the unique solution to the
initial value problem
\begin{equation}  \label{E:ivp4}
\mL \psi_\gl =\gl \psi_{\gl}\quad
 \text{with }  \psi_{\gl}(0)=p_{0} \text{ and } 
\frac{\gs(0)^{2}}{2}\psi'_{\gl}(0)=1-p_{0}.
\end{equation}

We summarize relevant results from Lemma 2.2, Theorem 3.4 
and Lemma 5.2 
of \cite{quasistat}.
\begin{Prop}
\label{P:omegat}
Assume the conditions (GB) and (LP).

\begin{itemize}
\item[(i)]
The eigenvalue
$\ulm$ is finite.
\item[(ii)]  
The semigroup $(P_{t})_{t \ge 0}$ has an asymptotic growth rate $\eta$.  
That is, for any compactly supported $\nu$, and any positive $s$,
\begin{equation}  \label{E:askil}
\ltf \frac{\langle P_{t+s}\indic,\nu\rangle}{\langle P_{t}\indic,\nu\rangle} = e^{\eta s}.
\end{equation}
\item[(iii)]
The following implications hold:
\[
\begin{split}
\ulm > \beta_\infty & \Longrightarrow \eta = \ulm > \beta_\infty \\
\ulm = \beta_\infty & \Longrightarrow \eta=\ulm=\beta_\infty \\
\ulm < \beta_\infty & \Longrightarrow \eta=\ulm \text{ or } \eta=\beta_\infty\\
\end{split}
\]
\item[(iv)]
If $\eta=\ulm \ne \beta_\infty$, then, for any compactly supported finite measure $\nu$,
\begin{equation}  \label{E:omegat}
\ltf \omega_{t}(\nu) =\int \frac{\psi_{\eta}(x)}{\psi_{\eta}(1)} \, d \nu(x),
\end{equation}
and, for any bounded test function $\varphi$, 
\begin{equation}  \label{E:quasistat}
\ltf \frac{\langle P_{t}\varphi,\nu\rangle}{\langle P_{t}\indic,\nu\rangle} 
= 
\frac{\izf \varphi(z)\phi_{\ulm}(z)dz}{\izf \phi_{\ulm}(z)dz}.
\end{equation}
\item[(v)]
If $\beta$ is non-constant and non-increasing and $p_0=1$
(that is, there is complete reflection at $0$),
then \eqref{E:quasistat} holds if and only if $\ulm > \beta_\infty$.
In particular, \eqref{E:quasistat} holds if $p_0=1$ and
$\beta$ is non-increasing and unbounded below (so that
$\beta_\infty=-\infty$).
\end{itemize}
\end{Prop}

\begin{proof}
The only part that is not copied directly from \cite{quasistat} is (v).
If $\ulm>\beta_\infty$, then $\eta = \ulm > \beta_\infty$ by (iii) and \eqref{E:quasistat}
holds by (iv). Suppose that \eqref{E:quasistat} holds but $\ulm \le \beta_\infty$.
Since there is no killing at $0$ ($p_{0}=1$), by \eqref{E:ivp}, the constant function is in the domain of $\mL$.  Thus, we may apply the Kolmogorov backward equation to see that by \eqref{E:askil},
\begin{align*}
\eta &=\frac{d\phantom{s}}{ds}\Bigr|_{s=0}\left( \ltf \frac{\langle P_{t+s}\indic,\nu\rangle}{\langle P_{t}\indic,\nu\rangle}\right)\\
&=\frac{d\phantom{s}}{ds}\Bigr|_{s=0} \left( \ltf \frac{\langle P_{t}(P_{s}\indic),\nu\rangle}{\langle P_{t}\indic,\nu\rangle}\right) \text{ by }\eqref{E:quasistat}\\
&=\frac{d\phantom{s}}{ds}\Bigr|_{s=0} \left( \frac{\izf P_{s}\indic(z)\phi_{\ulm}(z)dz}{\izf \phi_{\ulm}(z)dz}\right)\\
&=\frac{\izf \mL\indic(z)\phi_{\ulm}(z)dz}{\izf \phi_{\ulm}(z)dz}\\
&=\frac{\izf \beta(z)\phi_{\ulm}(z)dz}{\izf \phi_{\ulm}(z)dz}.
\end{align*}
Hence $\eta > \beta_\infty$ because $\beta$ is non-constant and $\phi_{\ulm}(z)$
is non-negative for all $z$ and strictly positive for $z$ close to $0$.
(In fact, $\phi_{\ulm}$ is strictly positive for all $z$).  However, 
this contradicts (iii).
\end{proof}

A simple consequence of Proposition \ref{P:omegat} is the following.

\begin{Cor}
\label{C:omegat}
For any compactly supported finite measure $\nu$ and any $\gep>0$, there are positive constants $0 < c_{\gep,\nu} \le C_{\gep,\nu} < \infty$ such that
if $0 \le t' \le t'' < \infty$, then
\begin{equation}  \label{E:cep}
c_{\gep,\nu}e^{(\eta-\gep)(t {''}-t')} \le \frac{\langle P_{t''}\indic,\nu\rangle}{\langle P_{t'}\indic,\nu\rangle}\le C_{\gep,\nu}e^{(\eta+\gep)(t''-t')}.
\end{equation}
\end{Cor}

\begin{proof}
By replacing $P_t$ by $e^{-\beta^* t}  P_t$, we may suppose that $(P_t)_{t \ge 0}$
is the semigroup of a killed diffusion, so that 
$\langle P_{t}\indic,\nu\rangle$ is non-increasing in $t$.

There is some integer $N$ such that if $N \le n$ is an
integer, then
\[
e^{\eta-\epsilon}
\le
\frac{\langle P_{n+1}\indic,\nu\rangle}
{\langle P_{n}\indic,\nu\rangle}
\le 
e^{\eta+\epsilon}.
\]
If $N \le n' \le n''$ are integers, then, since
\[
\frac{\langle P_{n''}\indic,\nu\rangle}
{\langle P_{n'}\indic,\nu\rangle}
=
\prod_{n=n'}^{n''-1} 
\frac{\langle P_{n+1}\indic,\nu\rangle}
{\langle P_{n}\indic,\nu\rangle},
\]
it follows that
\[
e^{(\eta-\epsilon)(n''-n')}
\le
\frac{\langle P_{n''}\indic,\nu\rangle}
{\langle P_{n'}\indic,\nu\rangle}
\le 
e^{(\eta+\epsilon)(n''-n')}.
\]

Note that if $n',n''$ are arbitrary non-negative integers, then
\[
\frac{\langle P_{n''}\indic,\nu\rangle}
{\langle P_{n'}\indic,\nu\rangle}
=
\frac{\langle P_{n''}\indic,\nu\rangle}
{\langle P_{n''\vee N}\indic,\nu\rangle} 
\frac{\langle P_{n' \vee N}\indic,\nu\rangle}
{\langle P_{n'}\indic,\nu\rangle}
\frac{\langle P_{n'' \vee N}\indic,\nu\rangle}
{\langle P_{n' \vee N}\indic,\nu\rangle}.
\]
On the one hand,
\begin{multline*}
\min_{m',m'' \le N}
\frac{\langle P_{m''}\indic,\nu\rangle}
{\langle P_{N}\indic,\nu\rangle} 
\frac{\langle P_{N}\indic,\nu\rangle}
{\langle P_{m'}\indic,\nu\rangle}
\le
\frac{\langle P_{n''}\indic,\nu\rangle}
{\langle P_{n''\vee N}\indic,\nu\rangle} 
\frac{\langle P_{n' \vee N}\indic,\nu\rangle}
{\langle P_{n'}\indic,\nu\rangle}\\
\le
\max_{m',m'' \le N}
\frac{\langle P_{m''}\indic,\nu\rangle}
{\langle P_{N}\indic,\nu\rangle} 
\frac{\langle P_{N}\indic,\nu\rangle}
{\langle P_{m'}\indic,\nu\rangle}.
\end{multline*}
On the other hand,
\[
e^{-N|\eta-\epsilon|}
e^{(\eta-\epsilon)(n''-n')}
\le
\frac{\langle P_{n'' \vee N}\indic,\nu\rangle}
{\langle P_{n' \vee N}\indic,\nu\rangle}
\le
e^{N|\eta+\epsilon|}
e^{(\eta+\epsilon)(n''-n')}.
\]
Thus the claimed inequality holds with suitable
constants for
$t',t''$ restricted to the non-negative
integers.  

For arbitrary $t'\le t''$ with 
$\lceil t' \rceil \le \lfloor t'' \rfloor$
we have
\[
\begin{split}
& c_{\gep,\nu} e^{-2|\eta-\gep|} e^{(\eta-\gep)(t''-t')}
\le
c_{\gep,\nu}e^{(\eta-\gep)(\lceil t'' \rceil - \lfloor t' \rfloor)} \\
& \quad \le
\frac{\langle P_{\lceil t'' \rceil}\indic,\nu\rangle}
{\langle P_{\lfloor t' \rfloor}\indic,\nu\rangle}
\le 
\frac{\langle P_{t''}\indic,\nu\rangle}
{\langle P_{t'}\indic,\nu\rangle}
\le
\frac{\langle P_{\lfloor t'' \rfloor}\indic,\nu\rangle}
{\langle P_{\lceil t' \rceil}\indic,\nu\rangle} \\
& \quad \le 
C_{\gep,\nu}e^{(\eta+\gep)(\lfloor t'' \rfloor - \lceil t' \rceil)}
\le e^{2|\eta+\gep|}
C_{\gep,\nu} e^{2|\eta+\gep|} e^{(\eta+\gep)(t''-t')}.\\
\end{split}
\]
and the result holds for such $t', t''$ by suitably adjusting the constants.

If $t'\le t''$ but 
$\lceil t' \rceil > \lfloor t'' \rfloor$
(so that $t''-t' \le 1$), then the observation
\[
\left(\frac{\langle P_{\lceil t' \rceil}\indic,\nu\rangle}
{\langle P_{\lfloor t'' \rfloor}\indic,\nu\rangle}\right)^{-1}
\le 
\frac{\langle P_{t''}\indic,\nu\rangle}
{\langle P_{t'}\indic,\nu\rangle}
\le
\left(\frac{\langle P_{\lfloor t' \rfloor}\indic,\nu\rangle}
{\langle P_{\lceil t'' \rceil}\indic,\nu\rangle}\right)^{-1}
\]
shows that a further adjustment of the
constants suffices to complete the proof.
\end{proof}

\subsection{Scaling limit}  \label{sec:rescaling}
The asymptotic behavior of Dawson-Watanabe superprocesses has been undertaken by Engl\"ander and Turaev \cite{ET02}, who applied general spectral theory to demonstrate, under certain formal conditions, the convergence in distribution
of the rescaled measure $e^{-\gl_{c}t}X_{t}$, where $\gl_{c}$ is the generalized principal eigenvalue.  Convergence in distribution was
improved to convergence in probability by Engl\"ander and Winter
\cite{EW06}.
We take a different route to proving a similar scaling limit, which has two advantages.
\begin{enumerate}
\item Our proof is more direct, and the conditions, when they hold, straightforward to check.
\item Our proof holds in some cases when the results of \cite{ET02} do not apply.  In particular, the scaling need not be exactly exponential.
\end{enumerate}
On the other hand, our approach is more restrictive, in  being applicable only to processes on $\R^{+}$.

Our proof depends on Dynkin's formulae \cite{eD91} for the moments of a Dawson-Watanabe superprocess.  For any bounded measurable $\varphi,\varphi':\R^{+}\to [-1,1]$, any starting distribution $\nu$, and any $0\le t\le t'$,
\begin{equation}  \label{E:dynkin}
\begin{split}
\E_{\nu} \left[\langle \varphi,X_{t}\rangle\right] &=\langle P_{t} \varphi,\nu\rangle,\\
\E_{\nu} \left[\langle \varphi,X_{t}\rangle\langle \varphi',X_{t'}\rangle\right] &=\langle P_{t} \varphi,\nu\rangle\langle P_{t'} \varphi',\nu\rangle \\ 
& \quad + \bigl\langle\int_{0}^{t} P_{s}
\left[\rho \cdot P_{t-s}  \varphi \cdot P_{t'-s}\varphi'\right]ds , \nu\bigr\rangle .
\end{split}
\end{equation}

Note by Proposition \ref{P:omegat} that in the following result the assumptions $\eta=\ulm>0$ and \eqref{E:quasistat} are implied by
the assumptions $\eta = \ulm > 0$ and $\ulm \ne \beta_\infty$.

\begin{Thm}
\label{T:rescaling}
Suppose that the conditions (LP) and (GB2) of Section \ref{sec:quasistat} hold.  Suppose further that
$\eta=\ulm>0$ and \eqref{E:quasistat} holds.
Then the rescaled random measure $\tX_{t}:=\langle P_{t}\indic,\nu\rangle^{-1} X_{t}$ converges in $L^{2}$ to a random multiple of the deterministic finite measure which has  density 
$\phi_{\ulm}$ with respect to Lebesgue measure.  That is, 
\begin{equation}  \label{E:testlim}
W :=\ltf \langle \indic,  \tX_{t}\rangle
\end{equation}
exists in $L^2$, and if $\varphi$ is any bounded test function,
\begin{equation}  \label{E:L2lim}
\ltf \E_{\nu}\left[\left\{\langle \varphi, \tX_{t}\rangle - 
W 
\frac{\int_0^\infty \varphi(z) \phi_{\ulm}(z) \, dz}
{\int_0^\infty \phi_{\ulm}(z) \, dz}
  \right\}^{2}\right]=0.
\end{equation}

The long-term growth rate of the
total mass of $X_t$ is
$\ltf t^{-1} \log\langle \indic, X_{t} \rangle = \ulm$.
\end{Thm}

\begin{proof}
From Proposition \ref{P:omegat}, we know that
$$
\ltf\E_{\nu} [\langle \varphi,\tX_{t}\rangle]
=\frac{\langle P_{t} \varphi,\nu\rangle}{\langle P_{t} \indic,\nu\rangle}=\frac{\izf \varphi(z) \phi_{\ulm}(z)dz}{\izf \phi_{\ulm}(z)dz}.
$$
Also,
\begin{align*}
& \langle P_{t}\indic, \nu \rangle^{-1} \langle P_{t'}\indic, \nu \rangle^{-1}
	\int_{0}^{t} P_{s}\left[\rho \cdot P_{t-s}\varphi \cdot P_{t'-s}\varphi'\right](x)ds \\
	& \quad =	\int_{0}^{\infty}\int \indic_{s\le t}\frac{\langle P_{t-s}\indic,\nu\rangle}{\langle P_{t}\indic,\nu\rangle}\frac{\langle P_{t'-s}\indic,\nu\rangle}{\langle P_{t'}\indic,\nu\rangle} \\
& \qquad \times	\rho(y)
\frac{\omega_{t-s}(y)}{\omega_{t-s}(\nu)}\frac{\omega_{t'-s}(y)}{\omega_{t'-s}(\nu)} \frac{P_{t-s}\varphi(y)}{P_{t-s}\indic(y)} \frac{P_{t'-s}\varphi'(y)}{P_{t'-s}\indic(y)}  P_{s}(x,dy) ds.
\end{align*}
For fixed $y$ the integrand converges as $t,t'\to \infty$ to
$$
\rho(y) \left(\frac{\psi_{\ulm}(y)}{\langle \psi_{\ulm}, \nu \rangle}\right)^{2}\left(\frac{\izf\varphi(z)\phi_{\ulm}(z)dz}{\izf \phi_{\ulm}(z)dz}\right)^{2} e^{-2\ulm s}.
$$

By \eqref{E:cep}, for any $t\le t'$ the integrand is bounded for any positive $\gep$ by
$$
\rho^*\rec{c_{\gep,\nu}} e^{-2s(\ulm-\gep)}\left(\sup_{\tau} \frac{\omega_{\tau}(y)}{\omega_{\tau}(\nu)}\right)^{2}
\le 
\rho^*\rec{c_{\gep,\nu}} e^{-2s(\ulm-\gep)}\omega_{*}(y,\nu)^{2},
$$
where $\rho^* := \sup_x \rho(x)$.
Condition (GB2) implies that this upper bound has finite integral for $\gep$ sufficiently small.  This allows us to apply the Dominated Convergence Theorem to conclude that
\begin{equation}\label{E:DCT}
\begin{split}
\lim_{t'\ge t\to\infty}\langle P_{t}&\indic,\nu\rangle^{-1}\langle P_{t'},\indic,\nu\rangle^{-1}
	\left\langle \int_{0}^{t} P_{s}\left[\rho \cdot P_{t-s}\varphi \cdot P_{t'-s}\varphi'\right]ds,\nu\right\rangle \\
	&=\left(\frac{\izf\varphi(z)\phi_{\ulm}(z)dz}{\langle \psi_{\ulm}, \nu\rangle\izf \phi_{\ulm}(z)dz}\right)^{2}\izf e^{-2\ulm s}\langle P_{s}\rho\psi_{\ulm}^2,\nu\rangle ds.
\end{split}
\end{equation}
That is, for any $\gep>0$, there is $T$ such that for all $t'\ge t\ge T$
$$
\left|\langle P_{t}\indic,\nu\rangle^{-1}\langle P_{t'},\indic,\nu\rangle^{-1}
	\left\langle \int_{0}^{t} P_{s}\left[\rho \cdot P_{t-s}\varphi \cdot P_{t'-s}\varphi'\right]ds,\nu\right\rangle-M(\nu,\varphi)\right|<\gep,
	$$
where $M(\nu,\varphi)$ is the right-hand side of \eqref{E:DCT}.

By \eqref{E:dynkin}, we may find $T_{\gep}$ such that for all $t'\ge t\ge T_{\gep}$,
\begin{align*}
\biggl|\E_{\nu} &[\langle \varphi,\tX_{t}\rangle\langle \varphi',\tX_{t'}\rangle] \\
&-K\left(\frac{\izf \varphi(z) \phi_{\ulm}(z)dz}{\izf \phi_{\ulm}(z)dz}\right)\left(\frac{\izf \varphi'(z) \phi_{\ulm}(z)dz}{\izf \phi_{\ulm}(z)dz}\right) \biggr|<\gep,
\end{align*}
where
$$
K:=1+\langle \psi_{\ulm}, \nu \rangle^{-2}
\izf e^{-2\ulm s}\langle P_{s}\rho\psi_{\ulm}^2,\nu\rangle ds.
$$
Then, for $t'\ge t\ge T_{\gep}$,
\begin{align*}
\E_{\nu} &\left[\left( \langle \varphi,\tX_{t}\rangle - \langle \varphi,\tX_{t'}\rangle \right)^{2}\right]\\
&=\E_{\nu} [\langle \varphi,\tX_{t}\rangle^{2}] 
+\E_{\nu} [\langle \varphi,\tX_{t'}\rangle^{2}] 
-2\E_{\nu} [\langle \varphi,\tX_{t}\rangle \langle \varphi,\tX_{t'}\rangle]\\
&< 4\gep.
\end{align*}
Thus $\langle \varphi,\tX_{t}\rangle$ is a Cauchy sequence in $L^{2}$
and converges to a limit.  In particular, there is a random finite measure $\tilde X_\infty$
such that $\tilde X_t \rightarrow \tilde X_\infty$ in probability in the topology of weak convergence
of finite measures on $\R^+$. 

It remains only to show that the limit is a multiple of the finite
measure with density $\phi_{\ulm}$.  For any bounded test function $\varphi$, let
\begin{align*}
A_{t}:=\langle \varphi,\tX_{t}\rangle - \langle \indic,\tX_{t}\rangle \langle \varphi,\phi_{\ulm}\rangle.
\end{align*}

By \eqref{E:dynkin},
\begin{align*}
\E[A^{2}_{t}]&=\left(\frac{\langle P_{t}\varphi,\nu\rangle}{\langle P_{t}\indic,\nu\rangle}\right)^{2}+\left\langle \int_{0}^{t}P_{s}\left[\rho \cdot \left(\frac{P_{t-s}\varphi}{\langle P_{t}\indic,\nu\rangle}\right)^{2}\right] ds,\nu \right\rangle\\
&\hspace*{1cm}+\langle \varphi,\phi_{\ulm}\rangle^{2}+\langle \varphi,\phi_{\ulm}\rangle^{2}\left\langle \int_{0}^{t}P_{s}\left[\rho \cdot \left( \frac{P_{t-s}\indic}{\langle P_{t}\indic,\nu\rangle}\right)^{2}\right] ds,\nu \right\rangle\\
&\hspace*{2cm}-2\langle \varphi,\phi_{\ulm}\rangle\frac{\langle P_{t}\varphi,\nu\rangle}{\langle P_{t}\indic,\nu\rangle}\\
&\hspace*{2cm}-2\langle \varphi,\phi_{\ulm}\rangle\left\langle \int_{0}^{t}P_{s}\left[\rho \cdot \frac{P_{t-s}\indic}{\langle P_{t}\indic,\nu\rangle}\frac{P_{t-s}\varphi}{\langle P_{t}\indic,\nu\rangle}\right] ds,\nu \right\rangle,
\end{align*}
which converges to $0$ as $t\to\infty$.
\end{proof}

\subsection{Optimal growth conditions}  \label{sec:highdif}
\begin{Thm}  \label{T:optimal}
Suppose $b$ is a fixed drift function, $\beta$ is a fixed non-increasing, non-constant
net reproduction function,
and $(\gs_m)$ is a sequence of diffusion functions.  Write $(P_{t}^{(m)})_{t \ge 0}$ for
the semigroup associated with $\gs_m$, $b$, and $\beta$, assuming complete
reflection at $0$.  Assume that \eqref{E:quasistat}
holds for each $m$. Denote the corresponding asymptotic growth rate
by $\ulm_m > \beta_\infty$.

Suppose that either of the following two conditions hold.
\begin{itemize}
\item[(a)]
\[
\inf_m \inf_{x\in\R^{+}}
\frac{b(x)}{\sigma_m(x)} - \sigma_m'(x) =: \gamma > -\infty
\]
and
\[
\lim_{m\to\infty}\inf_{x\in\R^{+}} \gs_m(x)=\infty,
\]
\item[(b)]
\[
\inf_{x\in\R^{+}}b(x) =: b_* > 0 
\]
and 
\[
\lim_{m\to\infty}\sup_{x\in\R^{+}} \gs_m(x)=0.
\]
\end{itemize}
Then $\lim_{m\to\infty}\ulm_m=\beta_\infty.$
\end{Thm}

\begin{proof}
The condition and the conclusion remain true if $\beta$ is changed by a constant, so we may assume, without loss of generality, that $\beta$ is non-positive.  This means that $P_{t}^{(m)}$ is the semigroup of a killed diffusion $\hat Y^{(m)}$.  Let $\tp^{(m)}$ be the killing time 
of $\hat Y^{(m)}$.  
Write $Y^{(m)}$ for the diffusion without killing that
corresponds to $\hat Y^{(m)}$.
Assume that $\beta_\infty > -\infty$.  The proofs
for $\beta_\infty = -\infty$ are similar and are left to the reader.
Since $\ulm_m > \beta_\infty$ always holds, it suffices in both cases
to establish that $\liminf_{m \to \infty} \ulm_m \le \beta_\infty$.

Consider first the case of condition (a).  
It follows from the assumption that $0$ is a regular
boundary point that the Liouville transformation
$F_m(x):=\int_{0}^{x} du/\gs_m(u)$
is finite for all $x \in \R^+$. 
The transformed diffusion $\hat V^{(m)} := F_m \circ \hat Y^{(m)}$ 
has diffusion function the constant $1$, drift function
\begin{equation}  \label{E:bF}
\frac{b \circ F_m^{-1}}{\gs \circ F_m^{-1}}-\oh\gs' \circ F_m^{-1},
\end{equation}
and killing rate function $-\beta \circ F_m^{-1}$.  
The asymptotic growth rate for $\hat V^{(m)}$ is also $\ulm_m$.
Write
$V^{(m)} := F_m \circ Y^{(m)}$ for the diffusion without
killing that corresponds to $\hat V^{(m)}$.

Since $\lim_{m\to\infty}\inf_{x\in\R^{+}} \gs_m(x)=\infty$, it follows
that $\lim_{m\to\infty} F_m^{-1}(x) = \infty$ for all $x>0$.  Hence,
given $\epsilon > 0$ there exists $M$ such that for $m \ge M$,
\[
\beta \circ F_m^{-1}(x) \le \beta_\epsilon(x),
\]
where
\[
\beta_\epsilon(x) :=
\begin{cases}
0, & 0 \le x < \epsilon, \\
(\beta_\infty + \epsilon) \frac{(x - \epsilon)}{\epsilon}, & \epsilon \le x < 2 \epsilon, \\
\beta_\infty + \epsilon, & x \ge 2 \epsilon.
\end{cases}
\]

Write $W$ for Brownian motion with drift $\gamma$ and complete
reflection at $0$.
By the comparison theorem for one-dimensional diffusions
and the assumption that $\beta$ is non-increasing,
\[
\begin{split}
\E\left[ \exp\left(\int_{0}^{t}\beta\circ F_m^{-1} (V^{(m)}_{s}) \, ds\right) \right]
& \le \E\left[ \exp\left(\int_{0}^{t}\beta\circ F_m^{-1}(W_{s}) \, ds\right) \right]\\
& \le \E\left[ \exp\left(\int_{0}^{t}\beta_\epsilon(W_{s}) \, ds\right) \right]\\
\end{split}
\]
if $V^{(m)}$ and $W$ have the same initial distribution.
Thus $\ulm_m \le \tilde \ulm_\epsilon$, where
\[
\tilde \ulm_\epsilon :=
\inf\ls \gl \, :\, \tilde \phi_{\gl, \epsilon}(x) \text{ is non-negative for all }x \rs
\]
for $\tilde \phi_{\gl, \epsilon}$ the unique solution to the
initial value problem
\begin{equation}
\label{E:eps_initial_value}
\frac{1}{2}\tilde \phi_{\gl, \epsilon}'' 
- \gamma \tilde \phi_{\gl, \epsilon}' +  \beta_\epsilon \tilde \phi_{\gl, \epsilon}
=\gl \tilde \phi_{\gl, \epsilon}\quad
 \text{with }  \tilde \phi_{\gl, \epsilon}(0)=1 \text{ and } 
\tilde \phi'_{\gl}(0) = \gamma.
\end{equation}

Solutions of the equation
\[
\frac{1}{2} f''(x) - \gamma f'(x) - r f(x) = \gl f(x)
\]
for some constant $r$ are linear combinations of the two functions
\[
\exp\left(\left(\gamma -\sqrt{\gamma^2+2 (\lambda + r)}\right) x\right)
\]
and
\[   
\exp\left(\left(\gamma+\sqrt{\gamma^2+2(\lambda + r)}\right) x\right).
\]
Solutions of the equation
\[
\frac{1}{2} f''(x) - \gamma f'(x) + (p - q x) f(x) = \gl f(x)
\]
for some constants $p$ and $q$ are linear combinations of the two functions
\[
e^{\gamma  x}
   \mathrm{Ai}\left(\frac{\gamma^2+2(\lambda -p + q
   x)}{2^{2/3} q^{2/3}}\right) 
\]
and
\[
e^{\gamma  x}
   \mathrm{Bi}\left(\frac{\gamma^2+2(\lambda - p+ q
   x)}{2^{2/3} q^{2/3}}\right),
\]
where $\mathrm{Ai}$ and $\mathrm{Bi}$ are the Airy functions
(see Section 10.4 of \cite{AS65}).
It follows that for fixed $\lambda$ the first and second
derivatives of $\tilde \phi_{\gl, \epsilon}$ are bounded on
$[0,\epsilon]$ and of orders at most $\epsilon^{-1/3}$ and 
$\epsilon^{-2/3}$, respectively, on $[\epsilon,2\epsilon]$.
Thus 
\begin{equation}  \label{E:eps}
\lim_{\epsilon \downarrow 0} 
\sup_{0 \le x \le 2 \epsilon}
|\tilde \phi_{\gl, \epsilon}(x) - 1| = 0 \text{ and }
\lim_{\epsilon \downarrow 0} 
\sup_{0 \le x \le 2 \epsilon}
|\tilde \phi_{\gl, \epsilon}'(x) - \gamma| = 0.
\end{equation}

Consider the initial value problem
\[
\frac{1}{2}\breve \phi_{\gl, c, g}'' 
- \gamma \breve \phi_{\gl, c, g}'' +  \beta_\infty \breve \phi_{\gl, c, g}
=\gl \breve \phi_{\gl, c, g}\quad
 \text{with }  \breve \phi_{\gl, c, g}(0)=c \text{ and } 
\breve \phi'_{{\gl, c, g}}(0) = g.
\]
Now
\[ 
\inf\ls \gl \, :\, \breve \phi_{\gl, 1, \gamma}(x) \text{ is non-negative for all }x \rs
\]
is simply the negative of the
asymptotic killing rate of reflected Brownian motion with drift 
$\gamma$ killed at constant rate $-\beta_\infty$, namely $\beta_\infty$ (this also follows
from explicit computation).  If $\lambda > \beta_\infty$, 
then an explicit
computation of the solution shows that $\breve \phi_{\gl, c, g}$
is non-negative for $c$ sufficiently close to $1$ and
$g$ sufficiently close to $\gamma$.  

Choose any $\lambda>\beta_{\infty}$.  On $(2\gep,\infty)$, the eigenfunction $\phi_{\gl, \epsilon}$ coincides with one of the functions $\breve \phi_{\gl, c, g}$.  From \eqref{E:eps} we know that by choosing $\gep$ sufficiently small, we can make $c$ as close as we like to 1 and $g$ as close as we like to $\gamma$; hence, $\phi_{\gl,\epsilon}$ is non-negative on $(2\gep,\infty)$.  The initial values of \eqref{E:eps_initial_value} show that it is also non-negative on $[0,2\gep]$ for $\gep$ sufficiently small.  Thus $\liminf_{\epsilon \downarrow 0} \tilde \ulm_\epsilon \le \lambda$; since $\gl$ was chosen arbitrary $\le \beta_{\infty}$, it follows
$\liminf_{\epsilon \downarrow 0} \tilde \ulm_\epsilon \le \beta_\infty$.  Combining this with \eqref{E:ulm}, we see that
and the proof for condition (a) is complete.

Now we consider the case of condition (b).  
Set
\[
\Sigma_m^2 =\sup_{x\in\R^{+}} \gs_m^{2}(x)
\]
and
\[
Z^{(m)}_{t}=Y^{(m)}_{t} - Y^{(m)}_0 - \int_{0}^{t}b(Y^{(m)}_{s})ds
- \ell_t^0(Y^{(m)}),
\]
where $\ell^0(Y^{(m)})$ is the local time at $0$ of $Y^{(m)}$.
Then $Z^{(m)}_{t}$ is a martingale with quadratic variation process bounded by $\Sigma_m^2 t$.  
Put
$$
Z^{(m)*}_{t}:=\sup_{0\le s\le t} \lv Z^{(m)}_{s}\rv.
$$
Because a continuous martingale is a time-change of Brownian motion, there is a universal constant $C$ such that
$$
\P\ls  Z^{(m)*}_{t}\ge \gep\rs \le C \exp(-\gep^{2}/2\Sigma_m^2 t).
$$

For any $\gd>0$,
\begin{align*}
\P\ls \tp^{(m)}>t\rs 
& = \E\left[ \exp\left(\int_{0}^{t}\beta(Y^{(m)}_{s}) \, ds\right) \right]\\
& \le \E\left[ \exp\left(\int_{\delta t}^{t}\beta(b_* s - b_{*}t\gd/2) \, ds\right) \right] + \P\ls Z^{(m)*}_{t}> b_{*}t\gd \rs \\
&\le \exp\left((1-\gd)t \sup_{x\ge b_{*}t\gd/2} \beta(x)\right)
  + \P\ls Z^{(m)*}_{t}> b_{*}t\gd/2 \rs\\
&\le \exp((1-\gd) \, \beta( b_{*}t\gd/2) \, t) + C \exp(-t(b_{*}\gd)^{2}/8\Sigma_m^2).
\end{align*}
(Recall we have assumed that $\beta$ is non-increasing.)
Since $\lim_{m \rightarrow \infty} \Sigma_m^2 = 0$ and 
$\lim_{x \rightarrow \infty} \beta(x) = \beta_\infty$, we may find $T$ such that
$$
\P\ls \tp^{(m)}>t\rs \le \exp((1-2\gd)\beta_\infty t)
$$
for all $t\ge T$ and all $m$ sufficiently large.  It follows that $\ulm_m \le (1-2\gd)\beta_\infty$
for $m$ sufficiently large.  
Thus $\liminf_{m \rightarrow \infty} \ulm_m  \le  \beta_\infty$, as required.
\end{proof}

\section*{Appendix: Convergence of the branching processes} \label{sec:modeldesc}

We show in this section that the sequence
of measure-valued processes $(X^n)$ converges to a Dawson-Watanabe
superprocess $X$ under suitable conditions. 
We first recall the definition of the limit $X$ as the
solution of a martingale problem.

For simplicity, we deal with the biologically most
relevant case where there is no killing at
$0$ in the damage and accumulation diffusion; that is,
$p_0 = 1$ in the boundary condition \eqref{E:ivp}, so that
the domain $\eu{D}(\eu{L})$ of the operator $\eu{L}$ 
consists of
twice differentiable functions $\phi$
that vanish at infinity, satisfy \eqref{E:ivp}
for $p_0=1$,
and are such that $\eu{L} \phi$ is continuous and vanishes
at infinity.
Similar arguments handle the other cases.

Suppose that the asymptotic rescaled branching rate
$\rho$ is bounded and continuous.  Recall that the
asymptotic net birth rate $\beta$ is continuous
and bounded above.  
Let $\mathcal{M}$ denote
the space of finite measures on $\R^+$ equipped
with the weak topology.  For each
$\nu \in \mathcal{M}$ there is a unique-in-law,
c\'adl\'ag, $\mathcal{M}$-valued process $X$ such that
$X_0 = \nu$ and
for every non-negative $\phi \in \eu{D}(\eu{L})$
\[
\exp(-\langle \phi, X_t \rangle) 
- \exp(-\langle \phi, \nu \rangle)
- \int_0^t \exp(-\langle \phi, X_s \rangle)
\langle -\eu{L} \phi   + \frac{1}{2}\rho \phi^2, X_s \rangle \, ds
\]
is a martingale.  Moreover, $X$ is  a Markov process with continuous paths.

Our convergence proof follows the proof of convergence of 
branching Markov processes to a Dawson-Watanabe superprocess, 
as found in Chapter 1 of \cite{aE00} or in Chapter 9 of \cite{EK86}.  
Because we  only wish to
indicate why a superprocess is a reasonable approximate model for
a large population of fissioning organisms, we don't strive for maximal
generality but instead impose  assumptions  that
enable us to carry out the proof with a minimal amount of technical detail.

Let $(\Pi_t)_{t \ge 0}$ be the 
Markovian semigroup generated  by
\[
\eu{G} \phi := \frac{1}{2} \sigma_{\mathrm{mot}}^2
\phi'' + b \phi'
\]
with boundary condition \eqref{E:ivp} for $p_0=1$.  
That is, the generator of $(\Pi_t)_{t \ge 0}$
is the operator $\eu{G}$ acting on the domain
$\eu{D}(\eu{G})$ consisting of
twice differentiable functions $\phi$
that vanish at infinity, satisfy \eqref{E:ivp}
for $p_0=1$ with $\sigma^2$ replaced by
$\sigma_{\mathrm{mot}}^2$,
and are such that $\eu{G} \phi$ is continuous and vanishes
at infinity.
Thus  $(\Pi_t)_{t \ge 0}$
is the Feller semigroup of the particle motion
for the measure-valued processes $X^n$.

Let $E$ denote the state space formed by taking the one-point
compactification of $\R^+$.  The semigroup $(\Pi_t)_{t \ge 0}$
can be extend to functions on $E$ in a standard manner.
The point
at infinity is a trap that is never reached
by particles starting elsewhere in $E$.
The domain of the extended semigroup is 
the span of $\eu{D}(\eu{G})$ and the constants 
(which we will also call $\eu{D}(\eu{G})$) and the generator
is extended by taking $\eu{G} \indic = 0$.
We also extend the measure-valued processes $X^n$
-- the point at infinity never collects any
mass if no mass starts there.

Set $\beta^n(x):= a^{n}(x)-k^{n}(x)$ and
$\rho^n(x):= n^{-1}(a^{n}(x)+k^{n}(x))$.
Suppose that $\beta^n$ and $\rho^n$ are bounded and continuous.
Given a bounded continuous function $g: \R^+ \rightarrow \R$,
put $\eu{J}_+^n g(x) := \int g(x+y) \, d\pi_x^n(y)$ and
$\eu{J}_-^n g(x) := \int g(x-y) \, d\pi_x^n(y)$

We have for $f \in \eu{D}(\eu{G})$ with $\inf_x f(x) > 0$ that
\[
\begin{split}
& \exp(\langle n \log f, X^n_t \rangle)
- \exp(\langle n \log f, X^n_0 \rangle) \\
& \quad - \int_0^t 
\exp(\langle n \log f, X^n_s \rangle) \\
& \qquad \times
\left\langle 
\frac{n}{f}
\left\{
 \eu{G} f + n \rho^n \left[\frac{a^n}{n\rho^n} \cdot \eu{J}_+^n f \cdot  \eu{J}_-^n f + \frac{k^n}{n\rho^n} - f \right]
\right\} 
, X^n_{s} 
\right \rangle
\, ds \\
\end{split}
\]
is a c\`adl\`ag martingale (cf. the discussion at the
beginning of Section 9.4 of \cite{EK86}).

If $X^n_0$ is a non-random measure 
$\nu \in \mathcal{M}$, then essentially
the same argument as in the proof of Lemma 9.4.1 of \cite{EK86}
shows that
\begin{equation}
\label{E:bounded_expect}
\E_\nu \left[\langle \indic, X_t^n \rangle\right]
\le
\langle \indic, \nu \rangle
\exp\left(t \sup_x \rho^n (x) \beta^n(x)\right)
\end{equation}
and
\begin{equation}
\label{E:tail_bound}
\P
\left\{
\sup_t
\, \langle \indic, X_t^n \rangle
\exp\left(- t \sup_x \rho^n(x) \beta^n(x)\right)
> z
\right\}
\le \frac{\langle \indic, \nu \rangle}{z}
\end{equation}
(the result in \cite{EK86} has the analogue of
$\sup_x |\rho^n (x) \beta^n(x)|$ in place
of $\sup_x \rho^n (x) \beta^n(x)$, but an examination of
the proof shows that it carries through with this constant).

In order to motivate our next set of assumptions, imagine just
for the moment that the jump distribution $\pi_x^n$ is
the distribution of a random variable $n^{-1/2} Y_x$ where
$Y_x$ has moments of all orders.  For $\phi$ sufficiently well-behaved
and $f_n := 1 - n^{-1} \phi$, a Taylor expansion shows that
\begin{equation}
\label{E:what_must_converge}
\frac{n}{f_n(x)}
\left\{
 \eu{G} f_n(x) + n \rho^n(x) \left[\frac{a^n(x)}{n\rho^n(x)} \cdot \eu{J}_+^n f_n(x) \cdot  \eu{J}_-^n f_n(x) + \frac{k^n(x)}{n\rho^n(x)} - f_n(x) \right]
\right\}
\end{equation}
is of the form
\[
\begin{split}
& -\Bigl\{
\frac{1}{2}\sigma_{\mathrm{mot}}^{2}(x) \phi''(x) + b(x) \phi'(x) + \beta^n(x) \phi(x) \\
& \quad + \frac{\rho^{n}(x)a^n(x)}{a^n(x)+k^n(x)} n \left(\int y^2 \, d\pi_x^n(y)\right) \phi''(x)
- \frac{\rho^n(x) a^n(x)}{a^n(x)+k^n(x)} \phi^2(x)
\Bigr\} \\
\end{split}
\]
plus lower order terms, and this converges pointwise to
\[
-\left\{
\frac{1}{2} \sigma^{2}(x) \phi''(x) + b(x) \phi'(x) + \beta(x) \phi(x)
- \frac{1}{2} \rho(x) \phi^2(x) \right\}
\]
under the assumptions \eqref{E:netbirth}, \eqref{E:branchvar},
and \eqref{E:segvar}.

With this informal observation in mind, assume that
$\eu{D}(\eu{L}) \subseteq \eu{D}(\eu{G})$ and that if 
$f_n := 1 - n^{-1} \phi$ with $\phi \in \eu{D}(\eu{L})$
and $\inf_x \phi(x) > 0$, then the function in
\eqref{E:what_must_converge} converges uniformly to 
$-\eu{L} \phi + \frac{1}{2} \rho \phi^2$.

Under this assumption, we can follow the proof of Theorem 9.4.3
in \cite{EK86} (using \eqref{E:tail_bound} to verify the
compact containment condition in the same manner that 
the similar bound (9.4.14) is used in \cite{EK86}) to establish
that if $X^n_0$ converges in distribution as a random measure on
$E$, then the process $X^n$ converges in distribution as
a c\`adl\`ag $E$-valued process to $X$.

\end{document}